%% file: SOCI_MCDM_Innovation.tex
\RequirePackage{rotating}
\RequirePackage{fix-cm}
\documentclass[smallextended]{svjour3}       % onecolumn (second format)
\smartqed  % flush right qed marks, e.g. at end of proof

\usepackage{graphicx,mathtools,booktabs,multicol,multirow,lscape}
\usepackage{amsmath,amssymb,stmaryrd}   
\usepackage{url}
\usepackage[table,xcdraw]{xcolor}
\usepackage{tabularx}
\usepackage{rotating}

\begin{document}

\title{Robust measurement of innovation performances in Europe with a hierarchy of interacting composite indicators}
%\subtitle{Do you have a subtitle?\\ If so, write it here}

\titlerunning{Robust measurement of innovation performances in Europe}        % if too long for running head

\author{S. Corrente, A. Garcia-Bernabeu, \\ S. Greco, T. Makkonen
}

\institute{Salvatore Corrente \at
              University of Catania, Italy \\
              \email{salvatore.corrente@unict.it}           %  \\
%             \emph{Present address:} of F. Author  %  if needed
           \and
           Ana Garcia-Bernabeu \at
              Universitat Polit\`ecnica de Val\`encia, Spain \\
              \email{angarber@upv.es}           %  \\
%             \emph{Present address:} of F. Author  %  if needed
           \and
					 Salvatore Greco \at
              University of Catania, Italy,  \\
							Univertisty of Portsmouth, UK, \\
              \email{salgreco@unict.it}
					\and
					Teemu Makkonen \at
					University of Eastern Finland, Finland \\
					\email{teemu.makkonen@uef.fi}
}

\date{Received: date / Accepted: date}
% The correct dates will be entered by the editor

\maketitle

\begin{abstract}
For long time the measurement of innovation has been in the forefront of policy makers' and researchers' agenda worldwide. Therefore, there is an ongoing debate about which indicators should be used to measure innovation. Recent approaches have favoured the use of composite innovation indicators. However, there is no consensus about the appropriate methodology to aggregate the varying dimensions of innovation into a single summary indicator. One of the best known examples of composite innovation indicators is the European Innovation Scoreboard (EIS). It is a relevant tool for benchmarking innovation in Europe. Still, the EIS lacks a proper scheme for weighting the included indicators according to their relative importance. In this context, we propose an appraisal methodology permitting to take into consideration the interaction of criteria and robustness concerns related to the elicitation of the weights assigned to the elementary indicators. With this aim, we apply the hierarchical-SMAA-Choquet integral approach. This integrated multicriteria decision making (MCDM) method helps the users to rank and benchmark countries' innovation performance taking into account the importance and interaction of criteria assigned by themselves, rather than equal weights or weights exogenously fixed by external experts.

% \PACS{PACS code1 \and PACS code2 \and more}
% \subclass{MSC code1 \and MSC code2 \and more}
\end{abstract}

\section{Introduction}
\label{intro}
According to recent scholarly debate, innovation is among, if not the main driver(s) of sustainable economic growth \cite{carayannis2018composite}. This has also been recognised by policy makers such as, for example, the European Commission (EC) that advocates policies for innovation-driven growth. The highlighted importance of innovation has induced a need for monitoring innovation performance across countries to understand and benchmark the success of policy measures in facilitating innovation and further economic growth. Some prime examples of these monitoring efforts are the Global Innovation Index (GII), which provides data and insights gathered from tracking innovation around the globe \cite{dutta2017}, and  the European Innovation Scoreboard (EIS) \cite{HollandersEsSadki2017} of which the latter is used here as an empirical case. Each year the EC publishes the annual EIS providing a comparative assessment of the research and innovation performance of the EU Member States and the relative strengths and weaknesses of their research and innovation systems. This indicator helps policy and decision makers to monitor, measure and benchmark the innovation performance of the EU, individual Member States, as well as associated countries and selected global competitors. It is a powerful tool for identifying areas in which the countries need to concentrate their efforts to boost their innovation performance \cite{foray2015assessment}. 

Since innovation activities are very complex, they are consequently challenging to measure \cite{holgersson2018towards}. There is no consensus about which indicators should be used to measure innovation. Initially, the use of individual indicators from national statistics such as the number of patents \cite{schmookler1950interpretation} or research and development (R\&D) expenditures \cite{griliches2007r} were the most commonly applied measures. However, starting from the 1990s, after pioneering works as \cite{hollenstein1996composite} who introduced the use of composite innovation indicators (CII's), scholars have started to favour and develop aggregate measures for innovation at the firm and geographical levels.  As such, the EIS also provides a composite indicator, the Summary Innovation Index (SII), in which the innovation performance is obtained by aggregating the indicators for each country into one single index. SII is composed of 27 indicators covering ten dimensions structured in four main blocks/pillars (Framework Conditions, Investments, Innovation Activities and Impacts). 

However, the application of such CII's for measuring (national) innovation performance and their utility in directing innovation policy has also been questioned (\cite{grupp2004indicators,schibany2008european,grupp2010review,adam2014measuring,edquist2015innovation,kozlowski2015innovation,hauser2017measuring}) mainly due to the problems related to the varying statistical and mathematical methods utilised for determining weights when aggregating the indicators into a CII \cite{carayannis2018composite}. Also, the latest EIS methodological report \cite{HollandersEsSadki2017b} claims the need for a new methodology in constructing the EIS: ``{\itshape it may be advisable for future refinements of the EIS to make use of a hierarchical structure in which indicators are first aggregated in dimension composites and subsequently, into group or overall averages}".

When constructing composite indicators, several critical concerns should be taken into account, specifically: weighting, aggregation of indicators, robustness, and the participation of experts in the construction of the composite indicators \cite{angilella2018robust}. Therefore, there is a broad consensus on the appropriateness of multiple criteria decision making (MCDM) methodologies to construct composite indicators \cite{saisana2002state,jacobs2004measuring,Nardo_2005} as they are highly suitable in multidimensional frameworks. Our approach is based on combining the recently proposed methodology called Multiple Criteria Hierarchy Process (MCHP) \cite{corrente2012multiple} with Stochastic Multicriteria Acceptability Analysis (SMAA)  \cite{lahdelma1998smaa}. This approach has been previously applied to evaluation problems concerning rankings of universities  \cite{angilella2016robust} and assessment of sustainable rural development \cite{angilella2018robust}.

The aim of this paper is to propose a CII based on the MCHP-Ch-SMAA methodology to fulfill the need for a hierarchical structure of the EIS and to overcome the criticism regarding weighting, aggregation and robustness in constructing composite indicators. Concerning the weighting problem and the aggregation of indicators, the proposed MCHP-Ch-SMAA approach employs an extension of the weighted sum based on the Choquet integral preference model \cite{choquet1953theory,grabisch1996application}, which takes into account the potential interactions between indicators. The CII is based on the judgements (or preferences) of experts representing the three helices of the Triple Helix model, that are, university, government and industry \cite{leydesdorff1998triple,etzkowitz2000dynamics}. To facilitate the case of responding and the ``readability" of  the results for decision makers, university, industry and government preferences with regards to the  dimensions of innovation are elicited from pairwise comparisons of indicators. Finally,  with respect to robustness, a probabilistic ranking is represented in terms of Rank Acceptability Indices (RAIs) for each country and Pairwise Winning Indices (PWIs) for each pair of countries. 

As a main contribution of the research, we propose the application of MCHP-Ch-SMAA to construct CII's overcoming the limitations of interaction of criteria and robustness concerns related to the elicitation of weights and their aggregation methodology. Trough the EIS application we also overcome the need of a proper scheme for weighting based on a hierarchical structure of dimensions and criteria, thus providing a measurement framework in which the preference information is not required at all the levels of the hierarchy reducing the cognitive effort of the decision maker. As to the challenge of estimating stakeholder’s preferences of innovation this approach provides a more in-depth analysis of the country’s innovation performance at the comprehensive level and for the specific macro-criterion incorporating the Triple Helix framework perspective. This approach is a useful tool to design and deploy policies and practices oriented towards specific Triple Helix agents as it permits to identify strengths and weaknesses for national innovation systems. 

The remainder of the paper is organized as follows. In Section 2,  the methodological basis regarding the issues of weighting, aggregation and robustness when constructing CII's are reviewed. Section 3 deals with the proposed methodology combining MCHP and SMAA. In Section 4, a case study related to the ranking of EU countries according to their innovation performance based on EIS criteria and incorporating Triple Helix agents point of view is presented. Conclusions and future lines of research are provided in Section 5.

\section{Weighting, aggregation and robustness in Composite Innovation Indicators (CII's)}
\label{sec:1}

During the past twenty years there has been an increasing interest from the research community in the methodological framework of constructing composite indicators, as they have gained increasing popularity as a benchmarking tool for national performances in a wide variety of fields ranging from socio-economic aspects to governance and environmental issues. However, and regardless of the decades of work on the topic, there still is no clear consensus on the best weighting and aggregation system as each method has its own strengths and limitations \cite{greco2018methodological}. Consequently, particularly problematic has been the assignment of appropriate weights for constructing the CIIs. %This issue is of extreme importance when considering the ability of CIIs to function as policy relevant indicators \cite{makkonen2013benchmarking, janger2017eu}. Therefore, and notwithstanding the popularity of composite indicators for country comparisons, their use has severe caveats: as stated by \cite{nardo2005tools}, in a benchmarking framework the weights and the aggregation methodologies can have a significant effect on the overall composite indicator and the ranking.  

% --------------------------
% --------------------------
% --------------------------
%\bibliographystyle{spphys}       % APS-like style for physics
%\bibliography{}   % name your BibTeX data base
%\bibliography{innovacion}
%\end{document}
% --------------------------
% --------------------------
% --------------------------

Basically, when addressing the weighting problem there are two main groups of approaches in which the weights can be obtained:

\begin{enumerate}

\item Objectively, based on statistical methods. Most composite indicators rely on equal weighting (EW) approaches in which all the variables are given the same weight. As such, applying equal weights is a simple but methodologically less robust approach to employing statistical tools to guide the weight calculations. Therefore an alternative is to utilize multivariate statistical approaches, such as, principal component analysis (PCA) \cite{AbdiWilliams2010}, in which the weights reflect the contribution of each indicator to the overall composite indicator. 

\item Subjectively, by means of participatory methods gathering stakeholders preferences to define the weighting scheme. In this group, the budget allocation approach (BAP), in which a panel of experts are given a fixed budget to be distributed over a number of dimensions, and  multi-attribute decision making methods such as the Analytic Hierarchy Process (AHP) \cite{saaty1977scaling} and conjoint analysis (CA) are among the most commonly applied methodologies for defining weights. In AHP, a complex decision problem is decomposed in a hierarchy of goal, criteria and alternatives  generating the weights according to the decision maker's pairwise comparisons of the criteria. Contrary, in the conjoint analysis the criteria weights are derived from the marginal rates of substitution of the overall utility function induced from preference information supplied by the user.
\end{enumerate} 

Another point of major concern is the aggregation methodology. While linear or additive aggregation prevails and it is the most widely used method compatible with all the previous weighting approaches, sometimes geometric aggregation is better suited  for the purpose of constructing composite indicators. However, in both cases  ``compensability''  appears \cite{paruolo2013ratings}, that is, the possibility of offsetting the shortfall in some dimension with a superior performance in another dimension. In multidimensional frameworks, when highly different dimensions should be aggregated, multicriteria decision making methodologies have been claimed as highly suitable alternatives for constructing composite indicators \cite{el2019building}. For example,  \cite{giannetti2009reliability} and \cite{arbolino2018towards} utilise elementary multicriteria methods such as simple additive weighting (SAW), to construct composite environmental indexes. Moreover, value and utility based methods \cite{keeney1976decision}, which include the Multi-Attribute Utility Theory (MAUT) and the Multi-Attribute Value Theory, have been utilized to construct a composite indicator, for example, by \cite{cracolici2009attractiveness} in their assessment on the attractiveness of tourism destinations, \cite{dantsis2010methodological} in their effort to measure the sustainability of agricultural plants and \cite{carayannis2016multilevel} who propose to evaluate the efficiency of national and regional innovation systems based on Data Envelopment Analysis (DEA) developed by \cite{charnes1978measuring}. The Benefit of the Doubt (BOD) methodology, rooted in DEA has been utilized to construct technology creation composite indicators \cite{cherchye2008creating}. The MACBETH (Measuring Attractiveness by a Categorical Based Evaluation Technique) approach introduced by \cite{e1994macbeth} has also been widely used for aggregating performance measurements (see e.g. \cite{cliville2007quantitative}). Concerning outranking methods based on comparisons between pairs of options, the most commonly used methods include ELECTRE (Eliminating and Choice Expressing Reality) \cite{FigueiraEtAl2013} and  PROMETHEE (Preference Ranking Optimization Method for Enrichment Evaluation) \cite{BransVincke1985}. For example, by using ELECTRE III, \cite{attardi2018non} develop a composite indicator for assessing the  environmental and social performances of urban and regional planning policies. \cite{petrovic2014electre} utilize the same methodology for benchmarking the performance of EU Member countries in achieving the key targets of EU's Digital Agenda, while \cite{antanasijevic2017differential} apply PROMETHEE II for assessing the sustainability performance of European countries. Additionally, the BOD methodology has been utilized to construct technology creation composite indicators \cite{cherchye2008creating}. 

 As shown above, the use of MCDM methodologies in constructing composite indicators has gained increasing popularity in recent years. However, there are only a few MCDM applications focusing on innovation and most of them are related to innovation planning and  technological roadmapping, except for the papers by \cite{paredes2014ranking} where they apply the Interactive and Multicriteria Decision Making (TODIM) (introduced by \cite{gomes1992todim}) to rank the performance of national innovation systems in the Iberian Peninsula and Latin America and, more recently, by \cite{carayannis2018composite}, who combine the AHP framework for setting priorities and the Technique for Order Preference by Similarity to Ideal Solution (TOPSIS) method \cite{hwang1981methods} for constructing CII. 

%Text with citations \cite{RefB} and \cite{RefJ}.

%%%%%%%%%%%%%%%%%%%%%%%%%%%%%%%%%%%%%%%%%%%%%%%%%%%%%%%%%%%%%%%%%%%%%%%%%%%%%%%%%%%%%%%%%%%%%%%%%%%%%%%%%%%%%%%%%%%%%%%%%%%%%%%%%%%%%%%%%%%%%%%%%%%%%%
\section{Multiple criteria hierarchy process (MCHP), Choquet integral preference model and Stochastic Multicriteria Acceptability Analysis (SMAA)}
\label{sec:2}
%%%%%%%%%%%%%%%%%%%%%%%%%%%%%%%%%%%%%%%%%%%%%%%%%%%%%%%%%%%%%%%%%%%%%%%%%%%%%%%%%%%%%%%%%%%%%%%%%%%%%%%%%%%%%%%%%%%%%%%%%%%%%%%%%%%%%%%%%%%%%%%%%%%%%%

\begin{figure}[h!]
 	
 	\includegraphics[width = 1\textwidth]{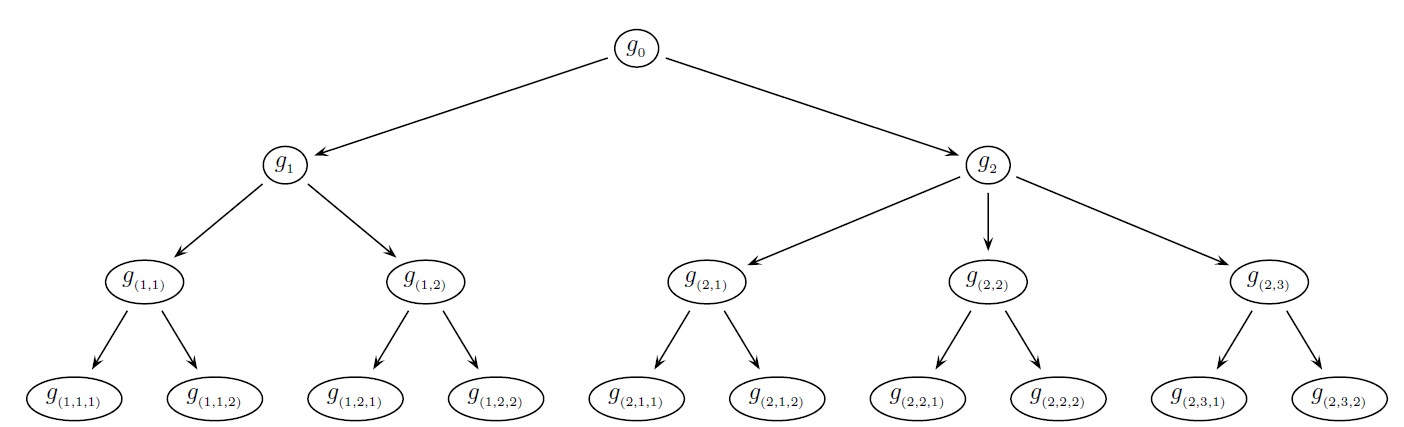}
 	\caption{Example on an hierarchy of criteria with 3 levels ($lev=3$), and 10 elementary criteria}
 	\label{Figura_Hierarchy}
 	
 \end{figure}

The Multiple Criteria Hierarchy Process (MCHP) \cite{corrente2012multiple} is a methodology recently introduced in Multiple Criteria Decision Aiding (MCDA) \cite{greco2016multiple} to deal with decision making problems in which the evaluation criteria are not at the same level but they are structured in a hierarchical way. This means that it is possible to define a root node $g_0$, being the objective of the problem, macro-criteria $g_1, \ldots , g_n$ descending from it, and so on, until the elementary criteria $g_{\mathbf{t}_1}, \ldots , g_{\mathbf{t}_n}$ being the criteria placed at the bottom of the hierarchy and on which the alternatives at hand are evaluated. See Figure \ref{Figura_Hierarchy} for an example of the hierarchy of criteria.

The advantage of taking into account the MCHP is two-fold: (i) it permits the Decision Maker (DM) to provide information on the alternatives not only at the comprehensive level (that is at the $g_0$ level), but also considering a particular criterion $g_\mathbf{r}$ in the hierarchy; (ii) it gives deeper information to the DM by defining a preference relation at each node of the hierarchy as well as at the comprehensive level. In this way, the DM can compare the alternatives not only globally, taking therefore into account all aspects simultaneously, but also focusing on the aspects which are relevant for him or her. 

In the following, we shall briefly recall the terminology that will be useful for the description of the methodology as well as for the discussion on the results of the considered application: 

\begin{itemize}
\item $A=\{a,b,\ldots\}$ denotes the set of alternatives at hand,
\item $lev$ denotes the number of levels in the hierarchy,
\item $g_0$ is the root criterion; it represents the main objective of the considered problem, 
\item $g_{\mathbf{r}}$ is a generic criterion in the hierarchy,
\item $G_{\mathbf{r}}^l$ is the set of subcriteria of $g_{\mathbf{r}}$ at the level $l$,
\item $G_{EL}=\{g_{\mathbf{t}_1}, \ldots, g_{\mathbf{t}_n}\}$ are the elementary criteria, that are the criteria at the lowest level of the hierarchy, while $EL$ is the set of their indices,
\item $E{(g_{\mathbf{r}})}$ is the set of the indices of the elementary criteria descending from $g_{\mathbf{r}}$.

\end{itemize}
To deal with any decision making problem, the evaluations of the alternatives on the considered criteria need to be aggregated by using one of the following aggregation methodologies that are the Multiple Attribute Value Theory (MAVT) \cite{keeney1976decision}, the outranking methods \cite{roy1996multicriteria} and the Dominance-Based Rough Set Approach (DRSA) \cite{greco2001rough}. In MAVT, one assigns a value $U(a)$ to each alternative $a \in A$ being representative of its goodness w.r.t. the problem at hand; outranking methods are based on the construction of a binary relation $S$, where $aSb$ means that ``$a$ is at least as good as $b$'' for all $a,b \in A$; finally, the DRSA build some ``if ...., then....'' decision rules expressed in a easily comprehensible language for the DM linking the alternatives' evaluations with the final recommendation. 

The Choquet integral preference model can be included under the MAVT \cite{choquet1953theory,grabisch1996application}. Indeed, it can be considered as a generalization of the weighted sum:

\begin{equation}
U(a)=\sum_{\mathbf{t}\in{EL}} w_{\mathbf{t}} \cdot g_{\mathbf{t}}(a)
\end{equation}

\noindent where $w_{\mathbf{t}}$ are the weights of criteria $g_{\mathbf{t}}$ such that $w_{\mathbf{t}} \geq 0$ for all $g_{\mathbf{t}}$  and $\sum_{\mathbf{t}\in{EL}} w_{\mathbf{t}} = 1$; moreover, $g_{\mathbf{t}}(a)$ is the evaluation of $a$ on  $g_{\mathbf{t}}$. 

Differently from the weighted sum, the use of the Choquet integral is based on a capacity $\mu:2^{|G_{EL}|}\rightarrow[0,1]$, being a set function that assigns a weight not only to each single criterion, but to all subsets of criteria $B\subseteq G_{EL}$ such that the monotonicity constraints ($\mu(B)\leq\mu(C)$ for all $B\subseteq C\subseteq G_{EL})$ and the normalization constraints ($\mu(\emptyset)=0$ and $\mu(G_{EL})=1$) are satisfied.

The main point of the Choquet integral preference model is that it is able to take into account the possible positive or negative interaction existing between criteria. Given $g_{\mathbf{t}},g_{\mathbf{t}_1}\in G_{EL}$, on one hand, we say that $g_{\mathbf{t}}$, $g_{\mathbf{t}_1}$ are positively interacting if the importance assigned to them together ($\mu (\{g_{\mathbf{t}},g_{\mathbf{t}_1}\})$) is greater than the sum of their importance when considered separately ($\mu (\{g_{\mathbf{t}}\}) + \mu (\{g_{\mathbf{t}_1}\})$), while, on the other hand, we say that $g_{\mathbf{t}}$ and $g_{\mathbf{t}_1}$ are negatively interacting if the importance assigned to them together is lower than the sum of their importance when considered separately. Of course, the same type of interactions can be defined for non-elementary criteria placed at the same level of the hierarchy of criteria.  

To make things easier, a M\"{o}bius transform of $\mu$ \cite{rota1964foundations} and $k$-additive capacities \cite{grabisch1996application}  are used in practical applications: 

\begin{itemize}
\itemsep=2ex
\item a M\"{o}bius transform of the capacity $\mu$  is a set function $m: 2^{G_{EL}}\rightarrow [0,1]$ such that  $\mu (B)=\sum_{C\subseteq B}m(C)$  for all $B\subseteq G_{EL}$,

\item  $\mu$ is called $k$-additive iff its M\"{o}bius transform is such that $m(B)=0$  for all $B\subseteq G_{EL}$ such that $\left|B\right|>k$ and there exists at least one $B\subseteq G_{EL}$, $|B|=k$, such that $m(B)=0$. In general, it is known that 2-additive capacities are able to perfectly represent all preferences provided by the DM \cite{mayag2011representation}. For this reason, in the following, we shall consider 2-additive capacities only and we shall briefly describe the 2-additive Choquet integral preference model.  

\end{itemize}

Considering 2-additive capacities, the monotonicity and normalization constraints above can be rewritten in the following way:

\begin{equation}
E^{Base}\left\{
\begin{array}{l}
    \left.\begin{array}{l}
         m(\{g_{\mathbf{t}}\})\geq 0\\[2ex]
         m(\{g_{\mathbf{t}}\}) + \sum\limits_{g_{\mathbf{t}_1}\in T}m (\{g_{\mathbf{t}},g_{\mathbf{t}_1}\})\geq 0 
    \end{array}\right\}
    \; 
    \begin{array}{l}
         \text{for all} \, g_{\mathbf{t}}\in G_{EL}\\[0ex]
         \text{and for all} \, T\subseteq G_{EL} \setminus \{g_{\mathbf{t}}\} 
    \end{array}\\[5ex]
    \sum\limits_{g_{\mathbf{t}}\in G_{EL}}m (\{g_{\mathbf{t}}\})+ 
        \sum\limits_{\{g_{\mathbf{t}},g_{\mathbf{t}_1}\}\subseteq G_{EL}} m (\{g_{\mathbf{t}},g_{\mathbf{t}_1}\})=1.
\end{array}\right.
\end{equation}

Therefore, for each $a \in A$ and for each criterion $g_{\mathbf{r}}$ in the hierarchy, the Choquet integral of $a$ on $g_{\mathbf{r}}$ is computed as 

\begin{equation}
Ch_{\mathbf{r}}(a)=\sum_{\mathbf{t} \in E(g_{\mathbf{r}})}m(\{g_{\mathbf{t}}\})\cdot g_{\mathbf{t}}(a) +\sum_{\{g_{\mathbf{t}}, g_{\mathbf{t}_1}\}\subseteq E(g_{\mathbf{r}})}m (\{g_{\mathbf{t}},g_{\mathbf{t}_1}\})\cdot \min \{g_{\mathbf{t}}(a),g_{\mathbf{t}_1}(a)\}.
\end{equation}

As already mentioned above, in applying the Choquet integral, a single value is assigned to each criterion as well as to all subsets of criteria. Therefore, the importance of a criterion is not dependent on itself only but also on its contribution to all coalitions of criteria. To take into account this aspect, the Shapley index \cite{shapley1953value}  and the Murofushi index \cite{murofushi1993techniques} are defined: 

\begin{itemize}
\item 	the Shapley index $\varphi_{\mathbf{r}}^l(\{g_{(\mathbf{r},w)}\})$, measures the importance of criterion $g_{(\mathbf{r},w)} \in G_{\mathbf{r}}^l$, that is considered as a subcriterion of $g_{\mathbf{r}}$ at the level $l$. Formally, it is computed as follows:

\begin{multline}
\varphi_{\mathbf{r}}^l(\{g_{(\mathbf{r},w)}\})= \left[
% To get the same size of both open and closed parenthesis
\vphantom{\sum_{\substack{
            \mathbf{t}_1\in E(g_{(\mathbf{r},w)})\\
            \mathbf{t}_2\in E(G_{\mathbf{t}}^l\setminus\{g_{{\mathbf{r}},w}\})}}
            \frac{m(\{g_{\mathbf{t}_1},g_{\mathbf{t}_2}\})}{2}}
\sum_{\mathbf{t}\in E(g_{(\mathbf{r},w)})} m(\{g_{\mathbf{t}}\}) + 
\sum_{\mathbf{t}_1,\mathbf{t}_2\in E(g_{(\mathbf{r},w)})}m(\{g_{\mathbf{t}_1},g_{\mathbf{t}_2}\}) \quad + \right.\\
\left. + \sum_{\substack{
            \mathbf{t}_1\in E(g_{(\mathbf{r},w)})\\
            \mathbf{t}_2\in E(G_{\mathbf{r}}^l\setminus\{g_{(\mathbf{r},w)}\})}}
     \frac{m(\{g_{\mathbf{t}_1},g_{\mathbf{t}_2}\})}{2} \right] \cdot \frac{1}{\mu(\{g_{\mathbf{t}}: \mathbf{t} \in Eg_{\mathbf{r}}\})}
\end{multline}
\noindent assuming that, of course, $\mu(\{g_{\mathbf{t}}: \mathbf{t}\in Eg_{\mathbf{r}}\})>0$;

\item the Murofushi and Soneda index $\varphi_{\mathbf{r}}^l(\{g_{(\mathbf{r},w_1)},g_{(\mathbf{r},w_2)}\})$, that measures the importance of the pair of criteria $\{g_{(\mathbf{r},w_1)},g_{(\mathbf{r},w_2)}\}\subseteq G_{\mathbf{r}}^l$ when considered as subcriteria of $g_\mathbf{r}$ at the level $l$. Formally, it is computed as follows: 

\begin{equation}
\varphi_{\mathbf{r}}^l(\{g_{(\mathbf{r},w_1)},g_{(\mathbf{r},w_2)}\})=
\sum_{\substack{
            \mathbf{t}_1\in E(g_{(\mathbf{r},w_1)})\\
            \mathbf{t}_2\in E(g_{(\mathbf{r},w_2)})}}
            m(\{g_{\mathbf{t}_1},g_{\mathbf{t}_2}\})\cdot \frac{1}{\mu(\{g_{\mathbf{t}}: \mathbf{t} \in E(g_{\mathbf{r}})\})}.
\end{equation}
\end{itemize}

A complete description of the extension of the Choquet integral to the MCHP can be found in \cite{angilella2016robust} where it has been introduced at first.

As it is evident from above, the application of the 2-additive Choquet integral preference model involves the knowledge of several parameters: considering the M\"{o}bius decomposition $m$, one value for each elementary criterion and one for each pair of elementary criteria. This is much less than the parameters to be elicited in case of a generic capacity not being 2-additive. Anyway, asking the DM to provide directly all these parameters is meaningless both for their huge number as well as for the difficult interpretation of their meaning. For this reason, to fix their values, an indirect elicitation procedure can be used. The DM is asked to provide information in terms of comparison between alternatives (for example, $a$  is preferred to $b$), comparison between criteria ($g_{(\mathbf{r},w_1)}$ is more important than $g_{(\mathbf{r},w_2)}$, with $g_{(\mathbf{r},w_1)},g_{(\mathbf{r},w_2)} \in G_{\mathbf{r}}^l$) or in terms of interactions between criteria ($g_{(\mathbf{r},w_1)}$ and $g_{(\mathbf{r},w_2)}$ are positively or negatively interacting). This preference information is therefore translated into inequality constraints (for example, the eventual preference of $a$ over $b$ on $g_{\mathbf{r}}$ is translated into the constraint $Ch_{\mathbf{r}}(a)\geq Ch_{\mathbf{r}}(b)+\varepsilon$ , while the positive interaction between  $g_{(\mathbf{r},w_1)}$ and $g_{(\mathbf{r},w_2)}$  is translated into the constraint $\varphi_{\mathbf{r}}^l(\{g_{(\mathbf{r},w_1)},g_{(\mathbf{r},w_2)}\}) \geq \varepsilon$; in both cases, $\varepsilon$ is an auxiliary variable used to transform the strict inequality constraints into weak inequality ones). Denoting by $E^{DM}$ the set of constraints translating the preferences provided by the DM, to check if there exists at least one instance of the preference model, that is one vector  $([m(\{g_{\mathbf{t}}\})]_{\mathbf{t} \in {EL}}, [m(\{g_{{\mathbf{t}}_1}, g_{{\mathbf{t}}_2}\})]_{{{\mathbf{t}}_1,{\mathbf{t}}_2} \in {EL}})$ for which all the technical constraints $(E^{Base})$ as well as all constraints translating the preferences provided by the DM ($E^{DM}$) are satisfied, one has to solve the following LP problem:

\begin{align}
&\varepsilon^{*}=\max \varepsilon, \text{  subject to}\\
& E=E^{Base} \cup E^{DM}.  \nonumber
\end{align}

\noindent If $E$ is feasible and $\varepsilon^{*}>0$, then there exists at least one instance of the preference model compatible with the preferences provided by the DM (briefly, a compatible model). In the opposite case, there does not exist any compatible model and the reason can be investigated by using one of the methods proposed in \cite{MousseauEtAl2003}.
In general, if there exists at least one compatible model, there exist many of them. Therefore, providing recommendations w.r.t. the problem at hand using only one of them can be considered arbitrary to some extent. Consequently, in the following, we shall describe the Stochastic Multicriteria Acceptability Analysis (SMAA) \cite{lahdelma1998smaa,tervonen2008survey}, methodology applied to the Choquet integral preference model that we have used in our application (for a recent extension of the SMAA methodology see \cite{greco2018stochastic,angilella2018robust,corrente2018evaluating}). 

SMAA provides the DM robust recommendations by considering not just one but the whole set of models compatible with the preferences given by the DM. Since the constraints in $E$ define a set composed by an infinite number of vectors, the application of SMAA begins with the sampling of several of them. We shall denote by $M$ the set composed of all sampled vectors. To each of these compatible vectors $M_j \in M$ corresponds an alternatives' ranking, where the rank position of the alternative $a$ w.r.t. criterion $g_{\mathbf{r}}$ obtained considering the vector of parameters $M_j$ is computed as
\begin{equation}
\textrm{rank}_{\mathbf{r}}(a,M_j)=1+\sum_{b\neq a} \rho(Ch_{\mathbf{r}}(b,M_j)>Ch_{\mathbf{r}}(a,M_j))
\end{equation}

\noindent where $\rho(false)=0$ and $\rho(true)=1$. Pay attention to the fact that in the definition of the rank function we used $Ch_{\mathbf{r}}(a,M_j)$ instead of $Ch_{\mathbf{r}}(a)$ as previously defined, just to underline that the computation of the Choquet integral is made by considering the parameters in the vector $M_j$. 

For each $a \in A$, for each criterion $g_{\mathbf{r}}$ and for each rank position $s=1, \dots , \left|A\right|$, is therefore possible to consider the set $M_{\mathbf{r}}^s(a) \subseteq M $  composed of the sampled compatible vectors giving to $a$ the position $s$ w.r.t. $g_{\mathbf{r}}$: 
\begin{equation} 
M_{\mathbf{r}}^s(a)=\{M_j \in M:\text{rank}_{\mathbf{r}}(a,M_j)=s\}.
\end{equation}

Analogously, for each pair of alternatives $a,b \in A$  and for each $g_{\mathbf{r}}$, it is possible to consider the set $M_{\mathbf{r}}(a,b)$ composed of all sampled compatible vectors for which $a$ is preferred to $b$ on $g_{\mathbf{r}}$: 
 \begin{equation} 
 M_{\mathbf{r}}(a,b)=\{M_j \in M:Ch_{\mathbf{r}}(a,M_j)> Ch_{\mathbf{r}}(b,M_j)\}.
 \end{equation}
The recommendations of SMAA are therefore given in statistical terms by computing the following indices:

\begin{itemize} 
\item \textbf{Rank Acceptability Index}, $b_{\mathbf{r}}^s(a)$: gives the frequency with which $a$ takes the $s-{th}$ position on criterion $g_{\mathbf{r}}$. It is obtained as

\begin{equation}
b_{\mathbf{r}}^s(a)=\frac{\lvert M_{\mathbf{r}}^s(a) \lvert}{\lvert M \lvert}.
\end{equation}

\item \textbf{Pairwise Winning Index,} $p_{\mathbf{r}}(a,b)$:  gives the frequency with which $a$ is preferred to $b$ on  $g_{\mathbf{r}}$. It is computed as 

\begin{equation}
p_{\mathbf{r}}^s(a,b)=\frac{\lvert M_{\mathbf{r}}(a,b) \lvert}{\lvert M \lvert}.
\end{equation}
\end{itemize}

On the basis of the Rank Acceptability Indices, for each alternative $a \in A$ and for each criterion $g_{\mathbf{r}}$ it is possible to compute the best and the worst positions got by $a$ on $g_{\mathbf{r}}$ as well as the most frequent ones \cite{angilella2018robust}. 

To conclude this section, we shall briefly summarize by means of the flow chart in \ref{Figura_Flow_Chart}. the main steps of the methodology that has been described and that will be applied to the Triple Helix innovation performance indicator:

\begin{figure}[h!]
 	
 	\includegraphics[width = 1\textwidth]{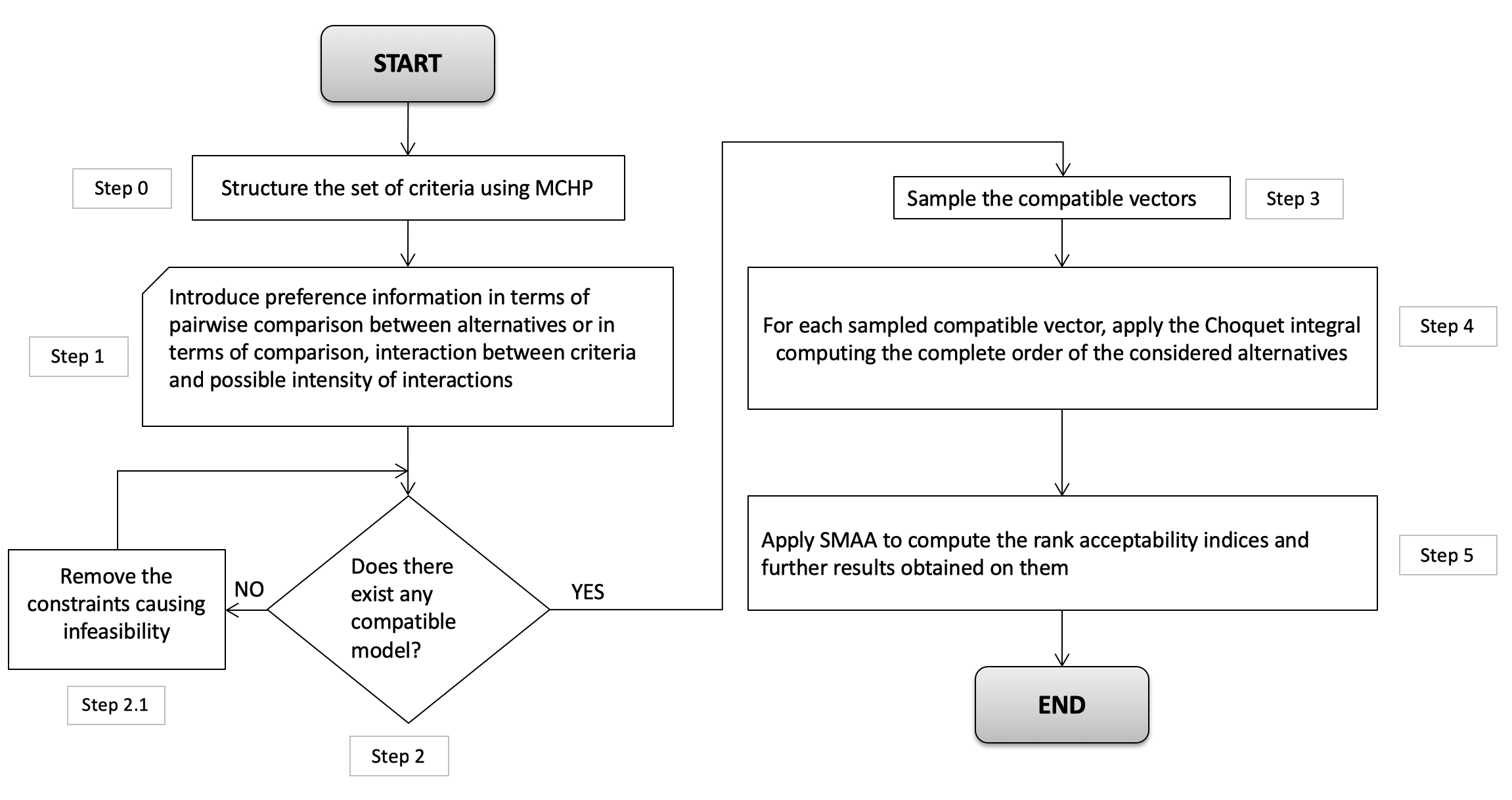}
 	\caption{Flowchart of the MCHP-Ch-SMAA approach}
 	\label{Figura_Flow_Chart}
 	
 \end{figure}

\begin{description}
\item \textbf{Step 0}: The criteria at hand are structured in a hierarchical way starting from the root until the elementary criteria; 
\item \textbf{Step 1}: Each DM is asked to provide his preference information that can be expressed in terms of comparison between alternatives, preferences between criteria, interaction between criteria, or intensity of interaction between criteria;
\item \textbf{Step 2}: Check if there exists at least one model compatible with the preferences provided by the DM by solving the LP (6). If this is not the case, check for the cause of the inconsistency and remove constraints causing the feasibility (Step 2.1). If there is at least one compatible model, pass to step 3;
\item \textbf{Step 3}: Sample several models (capacities in our case) compatible with the preferences provided by the DM in Step 1;
\item \textbf{Step 4}: For each compatible model sampled in Step 3, compute the Choquet integral of each alternative and, therefore, the consequent alternatives’ ranking;
\item \textbf{Step 5}: Apply the SMAA methodology computing for each alternative and for each position in the ranking, the rank acceptability index. On the basis of the obtained rank acceptability indices, compute, for each alternative:
\begin{enumerate}
\item Best and worst reachable positions,
\item The ranking positions presenting the highest rank acceptability indices that are, consequently, the most plausible positions for that alternative, 
\item The expected ranking obtained by aggregating the different rank acceptability indices as shown in the next section. 
\end{enumerate}
 
\end{description}

%%%%%%%%%%%%%%%%%%%%%%%%%%%%%%%%%%%%%%%%%%%%%%%%%%%%%%%%%%%%
\section{Application and results}
\label{sec:3}%%%
\begin{figure}[h!]
 	
 	\includegraphics[width = \textwidth]{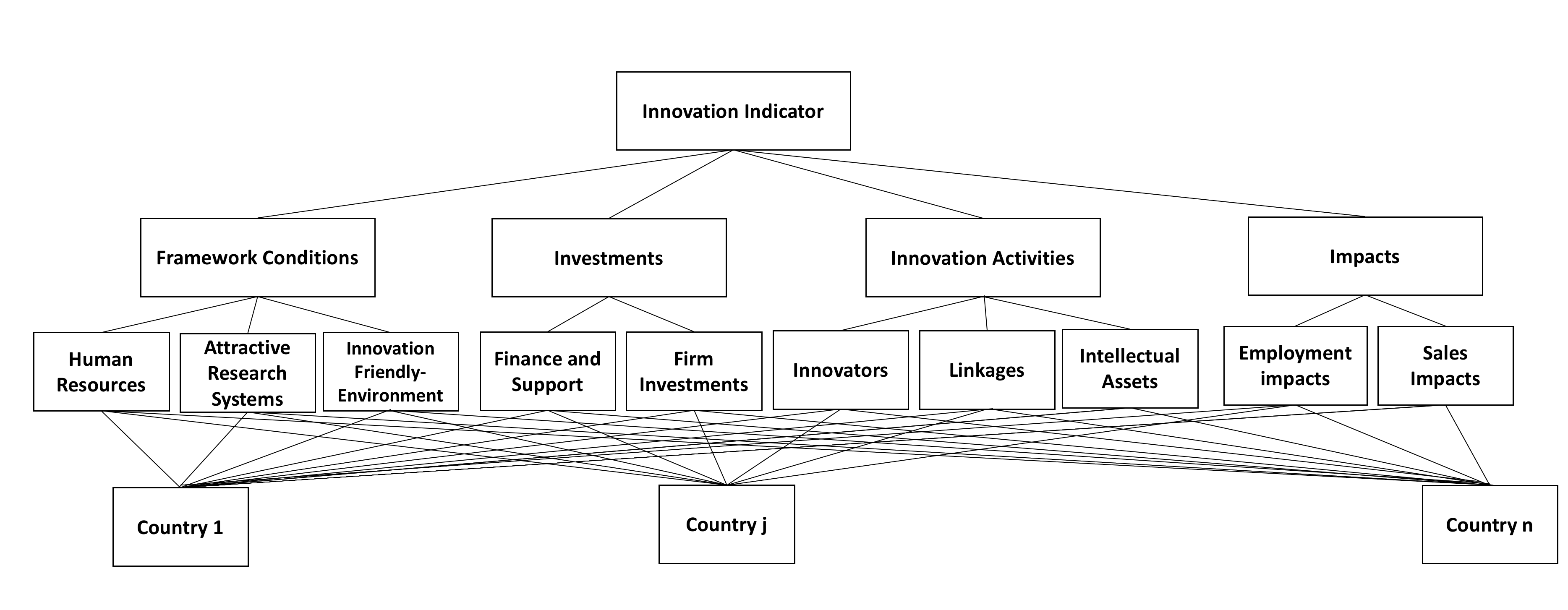}
 	\caption{Hierarchical structure framework for the EIS}
 	\label{Figura_AHP}
 	
 \end{figure}

%%%%%%%%%%%%%%%%%%%%%%%%%%%%%%%%%%%%%%%%%%%%%%%%%%%%%%%%%%%%
To show how the proposed methodology works, we develop a real world application in order to evaluate the innovation performance of 28 countries of the EU with respect to the SII criteria structured in a hierarchical way as shown in Figure \ref{Figura_AHP}.  The four macro-criteria are Framework Conditions (FC), Investments (IN), Innovation Activities (IA) and Impacts (IMP). They are further decomposed into more detailed criteria as follows:

\begin{description}

\item Macro-criterion (FC) is decomposed into: 

\begin{itemize}
 \item Human resources (HR)
 \item Attractive research systems (ARS)
  \item Innovation-friendly environment (IFE)
\end{itemize}

\item Macro-criterion (IN) is decomposed into: 

\begin{itemize}
 \item Finance and support (FS)
 \item Firm investments (FI)
  
\end{itemize}

\item Macro-criterion (IA) is decomposed into: 

\begin{itemize}
 \item Innovators (IT)
 \item Linkages (LIN)
 \item Intellectual assets (IAS)
  
\end{itemize}
\item Macro-criterion (IMP) is decomposed into: 
\begin{itemize}
 \item Employment impacts (EI)
 \item Sales effects (SE).
   
\end{itemize}

\end{description}

The description of each elementary criterion is given in Table \ref{tab:criteria-description}, while the evaluations of the 28 countries on these criteria are shown in Table \ref{tab:performances-countries} in the appendix. 
 
\begin{table}[htbp]
  \centering
  \caption{EIS description for the elementary criteria. Source: \cite{hollanders2016european}}
 \resizebox{\textwidth}{!}{\input{table_description_criteria}}
  \label{tab:criteria-description}%
\end{table}%

This research focuses on the perception of Triple Helix agents (University, Industry and Government) about the significance of criteria involved in the SII. These agents have different profiles to determine the degree of relative importance of criteria in order to obtain a composite indicator for innovation. To show the potential of our proposed methodology we have simulated the decision making process in order to obtain the information about preferences from a sample of Triple Helix agents. Through a consensus-driven decision-making process, three decision makers per each group agreed the preference information in the form of pairwise comparisons related to importance and interaction of the four macro criteria as well as for the elementary criteria. A pool of experts from different European countries with solid background in knowledge and professional experience concerning innovation has been consulted aiming at reflecting different views of the relative importance of criteria:

\begin{itemize}
\item University Experts (DMU): academics expert in the field of innovation from the United Kingdom, technical staff of a University Technology Transfer Office from Spain, and academic expert in the field of EU innovation policies from Spain.
\item Industry Experts (DMI): business manager in the field of international business from Belgium, chief innovation officer from Spain, and project manager from Italy.
\item Government Experts (DMG): R\&D program manager from Germany, the director of a Technological Institute from Spain, and deputy director of a patent and trademark office from Poland.
\end{itemize}

The DMU group specified the following preference information on the considered macro-criteria as follows. 

\begin{itemize}
\item FC and IN are more important than IA and IMP; in turn FC is equally important than IN, and IA is more important than IMP.
\item With respect to FC, HR and ARS are more important than IFE.
\item With respect to IN, FS is more important than FI.
\item With respect to IA, LIN is more important than IT and IAS.
\item With respect to IMP, EI is more important than SE.
\item FC and IN are positively interacting.
\item IN and IA are positively interacting.
\item IA and IMP are positively interacting. 
\item The interaction between IN and IA is greater than the interaction between FC and IN, and between IA and IMP.
\item The interaction between IA and IMP is greater than the interaction between FC and IN.
\item With respect to FC, HR and ARS are positively interacting.
\item With respect to IA, IAS and LIN are positively interacting.
\item IAS and IT are negatively interacting.
\end{itemize}

For DMI group, the preference information on the considered macro-criteria was:

\begin{itemize}
\item IN  is more important than IA, that in turn is more important than  IMP, that in turn, is more important than FC. 
\item With respect to FC, ARS is more important than IFE, which is more important than HR.
\item With respect to IN, FS is more important than FI.
\item With respect to IA, LIN is more important than IT, that in turn, is more important than IAS.
\item With respect to IMP, SE is more important than EI.
\item FC and IN are positively interacting.
\item IN and IA are positively interacting.
\item IA and IMP are positively interacting. 
\item The interaction between IN and IA is greater than the interaction between FC and IN, and  between IA and IMP.
\item The interaction between IA and IMP is greater than the interaction between FC and IN.
\item With respect to FC, HR and ARS are positively interacting.
\item With respect to IA, IAS and LIN are positively interacting.
\item IAS and IT are negatively interacting.
\end{itemize}

Finally, for the DMG group the preference information on the considered macro-criteria was:

\begin{itemize}
\item IMP is more important than FC, that in turn, is more important than IN, that in turn, is more important than IA. 
\item With respect to FC, ARS is more important than IFE, that in turn, is more important than HR.
\item With respect to IN, FS is more important than FI.
\item With respect to IA, LIN is more important than IAS, that in turn, is more important than IT.
\item With respect to IMP, SE is more important than EI.
\item FC and IN are positively interacting.
\item IN and IMP are positively interacting.
\item IA and IMP are positively interacting. 
\item The interaction between IN and IMP is greater than the interaction between FC and IN, and IA and IMP.
\item The interaction between IA and IMP is greater than the interaction between FC and IN.
\item With respect to FC, HR and ARS are positively interacting.
\item With respect to IA, IAS and LIN are positively interacting.
\item IAS and IT are negatively interacting.
\end{itemize}

The use of the Choquet integral preference model implies that all evaluations are expressed on the same scale. For this reason, before applying it, we performed a normalization of the countries performances proposed in \cite{GrecoEtAl2018} and composed of the following steps: 
\begin{enumerate}
\item First step: for each elementary criterion $g_{\mathbf{t}}$, compute the mean $M_{\mathbf{t}}$ and the standard deviation $s_{\mathbf{t}}$ of the countries performances on that criterion: 
$$
M_{\mathbf{t}}=\frac{1}{|A|}\displaystyle\sum_{a\in A}g_{\mathbf{t}}(a), \qquad s_{\mathbf{t}}=\sqrt{\frac{\displaystyle\sum_{a\in A}\left(g_{\mathbf{t}}(a)-M_{\mathbf{t}}\right)^2}{|A|}}
$$
\noindent where $A$ denotes the set of countries and $g_{\mathbf{t}}(a)$ is the performance of country $a$ on $g_{\mathbf{t}}$;
\item Second step: for each $a$ and for each $g_{\mathbf{t}}$, the $z$-score $g^{z}_{\mathbf{t}}(a)$ is computed: $\displaystyle g^{z}_{\mathbf{t}}(a)=\frac{g_{\mathbf{t}}(a)-M_{\mathbf{t}}}{s_{\mathbf{t}}}$;
\item Third step: for each $a$ and for each $g_{\mathbf{t}}$, the normalized evaluation $\overline{g}_{\mathbf{t}}(a)$ is therefore obtained as
$$
\overline{g}_{\mathbf{t}}(a)=
\left\{
\begin{array}{lll}
0 & \mbox{if} & g_{\mathbf{t}}(a)\leq M_{\mathbf{t}}-3s_{\mathbf{t}},\\[0,2mm]
0.5+\frac{g_{\mathbf{t}}^{z}(a)}{6}, & \mbox{if} & M_{\mathbf{t}}-3s_{\mathbf{t}}<g_{\mathbf{t}}(a)<M_{\mathbf{t}}+3s_{\mathbf{t}}, \\[0,2mm]
1, & \mbox{if} & g_{\mathbf{t}}(a)\geq M_{\mathbf{t}}+3s_{\mathbf{t}}
\end{array}
\right.
$$
\noindent if $g_{\mathbf{t}}$ has an increasing direction of preference and 
$$
\overline{g}_{\mathbf{t}}(a)=
\left\{
\begin{array}{lll}
0 & \mbox{if} & g_{\mathbf{t}}(a)\geq M_{\mathbf{t}}+3s_{\mathbf{t}},\\[0,2mm]
0.5-\frac{g_{\mathbf{t}}^{z}(a)}{6}, & \mbox{if} & M_{\mathbf{t}}-3s_{\mathbf{t}}<g_{\mathbf{t}}(a)<M_{\mathbf{t}}+3s_{\mathbf{t}}, \\[0,2mm]
1, & \mbox{if} & g_{\mathbf{t}}(a)\leq M_{\mathbf{t}}-3s_{\mathbf{t}}
\end{array}
\right.
$$
\noindent if, instead, $g_{\mathbf{t}}$ has a decreasing direction of preference\footnote{On one hand, a criterion $g_{\mathbf{t}}$ has an increasing direction of preference if the more $g_{\mathbf{t}}(a)$, the better is $a$ on $g_{\mathbf{t}}$, while, on the other hand, a criterion $g_{\mathbf{t}}$ has a decreasing direction of preference if the less $g_{\mathbf{t}}(a)$, the better is $a$ on $g_{\mathbf{t}}$.}.
\end{enumerate}  
The normalized evaluations on which the methodology is applied are shown in Table \ref{tab:normalized_performances-countries} in the appendix.
 
By applying the MCHP-Ch-SMAA approach to the Triple Helix agents at the comprehensive level and for the four macro-criteria, the rank acceptability indices (RAI) for the best and worst performers as well as the three highest rank acceptability (most frequent positions) indices showing which are the most likely position for a country based on the appreciation of each actor's capabilities and needs are calculated. Tables \ref{tab:table_RAI_DMU}-\ref{tab:table_RAI_DMG} presents the results concerning the top five and to the bottom three countries. The following observations can be made based on the results:

\begin{itemize}
\itemsep=2ex
\item With respect to DMU, the top five innovative countries are Sweden, Denmark, Finland, Netherlands and Germany and the last three positions are for Croatia, Bulgaria and Romania. At the macro-criterion level, it is important to note that the rank position for Denmark and Finland significantly varies for the Impact (IMP) dimension. Denmark reached as best position the 7th even if the rank acceptability index is basically zero (0.70\%); moreover it reaches most frequently the 12th position  with 45.04\%. Finland ranks the 10th position  with a frequency of 4.11\% for the best and 20.69\% for the first most frequent position, being the 13th. At the same time, Germany also gets worse positions at the Framework Conditions (FC) being in the 12th position presenting a rank acceptability index of 10.19\%. Concerning the last three countries the case of Croatia is worth pointing out,  in which its position with respect to macro-criterion investments (IN) moved up to 16th position, even if its frequency is negligible (0.25\%) for the best position. 

\item With respect to DMI, Sweden, Finland, Netherlands, United Kingdom and Denmark can be considered the most innovative countries whereas, Poland, Bulgaria and Romania are in the worst positions. In particular, Finland appears to be quite unstable with respect to macro-criterion Impacts (IMP), although its highest rank acceptability index for the 10th position has a marginal frequency and the most frequent positions range between 17th (24.52\%) and 18th position (21.07\%). From the industry perspective, in Denmark the dimension corresponding to framework conditions (FC) is highlighted with a rank acceptability index of 98.29\%.  Among the last three countries, quite stable results are presented at all levels.

\item With respect to DMG, the top five countries are Sweden, United Kingdom, Denmark, Netherlands, and Ireland. Croatia, Bulgaria and Romania are placed at the bottom of the ranking. One can observe that Ireland improves its position with respect to  DMI and DMG, and moreover, its rank acceptability index for the first position in the macro-criterion impacts (IMP) is quite good (86.33\%). Regarding the last positions, while Bulgaria and Romania present the highest rank acceptability indices for the 27th and 28th position (100\%), respectively,  Croatia has a frequency of 22.04\%  of being in the 25th position. Furthermore, Croatia has risen to the 7th place in the Investments (IN) with a frequency of 0.11\%  and to 15th first most frequent position with 52.91\%.
\end{itemize}

\begin{table}[htbp]
	\centering
	\caption{Rank Acceptability Indices, best and worst positions and three highest rank acceptability indices from DMU at the comprehensive level and for the four macro-criteria}
	\resizebox{0.85\textwidth}{!}{
		\input{table_RAI_DMU}
		}
	\label{tab:table_RAI_DMU}%
\end{table}

\begin{table}[htbp]
	\centering
	\caption{Rank Acceptability Indices, best and worst positions and three highest rank acceptability indices from DMI at the comprehensive level and for the four macro-criteria}
	\resizebox{0.85\textwidth}{!}{

\input{table_RAI_DMI}

		}
	\label{tab:table_RAI_DMI}%
\end{table}

\begin{table}[htbp]
	\centering
	\caption{Rank Acceptability Indices, best and worst positions and three highest rank acceptability indices from DMG at the comprehensive level and for the four macro-criteria}
	\resizebox{0.85\textwidth}{!}{
		\input{table_RAI_DMG}
		}
	\label{tab:table_RAI_DMG}%
\end{table}

As explained in \cite{KadzinskiMichalski2016}, the expected ranking is computed as a summary of the rank acceptability indices of each country. Analogously, the expected ranking of each country at comprehensive level is computed and results are shown in Table \ref{tab:table_summary}. For each $a\in A$, and for each criterion $g_{\mathbf{r}}$ in the hierarchy, the expected ranking of $a$ on $g_{\mathbf{r}}$ is computed as follows: 

$$
E_{\mathbf{r}}(a)=-\displaystyle\sum_{s=1}^{|A|}s\cdot b_{\mathbf{r}}^{s}(a)  
$$

\noindent On the basis of the values $E_{\mathbf{r}}(a)$, the countries are therefore ordered from the best to the worst on each criterion $g_{\mathbf{r}}$.

 \begin{figure}[h!] 	
 	\includegraphics[width = \textwidth]{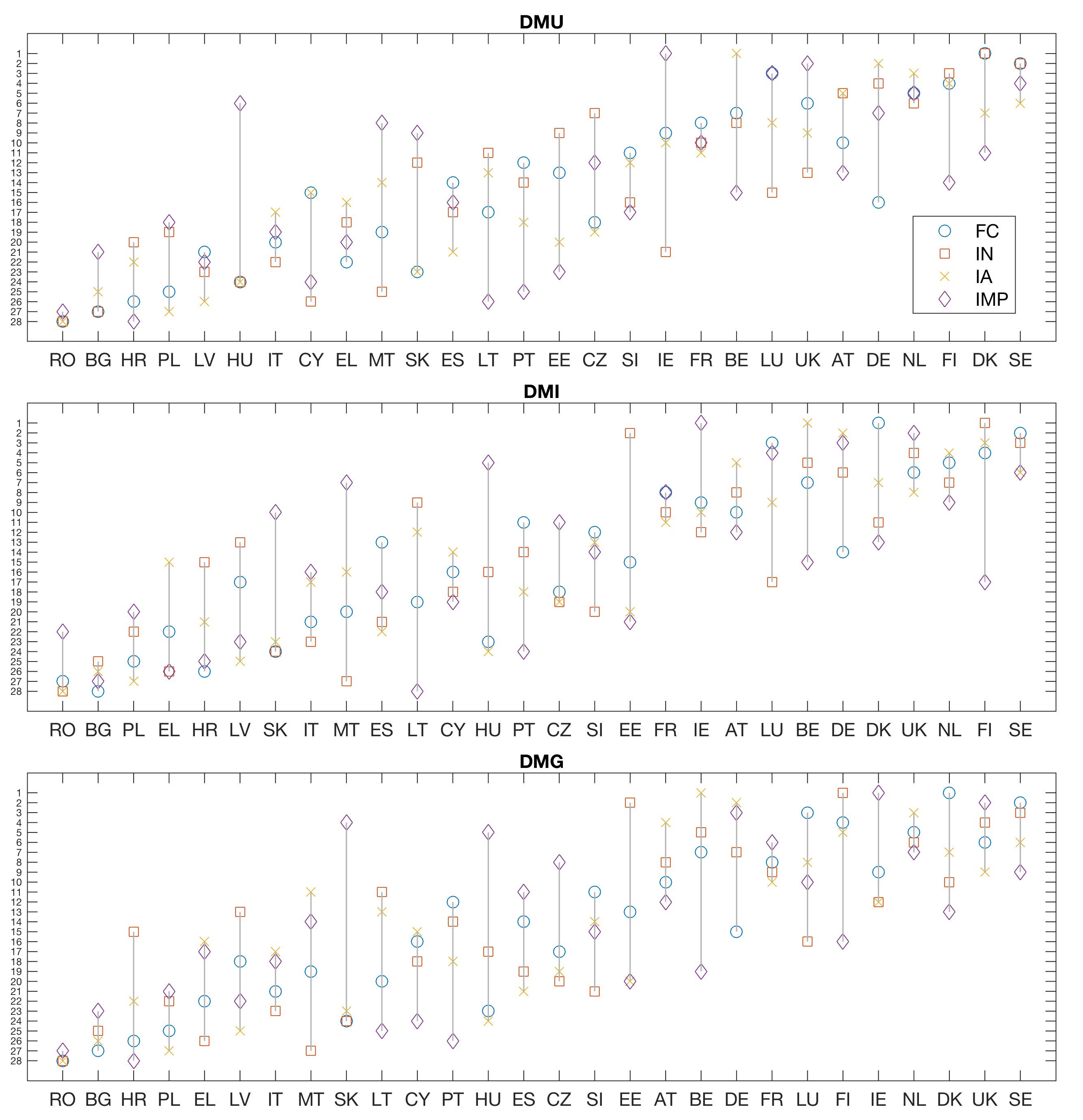}
 	\caption{EU countries innovative performance expected rankings for the Triple Helix agents}
 	\label{Figura_DMU_DMI_DMG} 	
 \end{figure}

Figure \ref{Figura_DMU_DMI_DMG} plots the expected rankings of the countries at the comprehensive level in the X-axis as well as the expected rankings at the macro-criteria level in the Y-axis per each Triple Helix actor. In this figure, top tier countries appear on the top right-hand corner for each Triple Helix agent and also maintain leading positions at the macro-criteria level. When analysing Figure \ref{Figura_DMU_DMI_DMG} we should draw attention to some countries. Looking at the comprehensive level (X-axis), Finland (FI), one of the nations in the top five, lies in the third position according to university decision makers (DMU), the second from the industry (DMI) perspective but only sixth from the governments ranking (DMG). When looking at the  macro-criteria level (Y-axis), we can observe  that the macro-criterion investments (IN) are the best valued and impacts (IMP) has the worst position for all the actors.

Another example is Ireland (IE) which has significant positive discrepancies in the government (DMG) ranking (fifth position) while it has a worse position (eleventh and tenth) according to university (DMU) and industry (DMI) preferences (X-axis). Attending at the macro-criteria level (Y-axis) we can observe some similarities across all the Triple Helix actors: impacts (IMP) were considered the most and investments (IN) the least valued dimension by all the groups of experts but mainly by university point of view.

To better understand the potentialities of our approach, in  Table \ref{tab:table_summary} we reported also the global country ranking provided by the EIS. Looking at the results,  it seems there is a consensus among the Triple Helix agents that Sweden is the most innovative country while Bulgaria and Romania get the worst positions in all rankings. There is a small difference in the positions of Croatia, Czech Republic, Italy, Latvia, Luxembourg,  Netherlands, Poland, Portugal and Slovenia. However, there is a significant deviation of at least three ranking positions in Belgium, Denmark, Germany, Estonia, Ireland, Greece, Spain, France, Cyprus, Lithuania, Hungary, Malta, Austria, Slovakia, Finland, and United Kingdom (see Table \ref{tab:table_discrepancies}).

\begin{table}[htbp]
	\centering
	\caption{Expected ranking of countries from the Triple Helix agents at the comprehensive level}
		\footnotesize
		\input{table_summary}
		%}
	\label{tab:table_summary}%
\end{table}

\begin{table}[htbp]
	\centering
	\caption{Significant discrepancies (at least three rank positions) from some of the Triple Helix Agent expected ranking and the EIS ranking}
		\footnotesize
		\input{table_discrepancies}
		%}
	\label{tab:table_discrepancies}%
\end{table}

The complete results of the application of MCHP-Ch-SMAA methodology to the 28 countries at the comprehensive level and to the four macro-criteria  are provided for the interested reader by downloading the following folder:

{\noindent\scriptsize\url{www.antoniocorrente.it/wwwsn/images/allegati_articoli/Supplementary%20Material.zip}}.

%%%%%%%%%%%%%%%%%%%%%%%%
\section{Concluding remarks and future research}
\label{sec:conclusions}%%%
%%%%%%%%%%%%%%%%%%%%%%%%
Promoting innovation is considered a key policy instrument to enhance competitiveness. To this end, it is also important to monitor national innovation performances. Recent approaches for assessing the innovation of countries have adopted composite indicators. In this paper, we firstly reviewed the state of the art and outlined recent development on CIIs. One of the main criticism made in the construction of composite indicators concerns the issues of weighting and aggregation as well as the necessity to take into account a hierarchical structure of criteria. In the weighting stage, objective approaches such as ``equal weighting" are one of the most widely used methods to avoid inconsistencies or subjectivity. However equal weighting is often considered as unrealistic. Therefore, to overcome the limitations of the existing EIS approach based on ``equal weighting'' we propose a methodology that incorporates different preferences of the Triple Helix agents (university, industry and government) by utilizing a multicriteria decision framework combining the Multiple Criteria Hierarchy Process (MCHP) with the Stochastic Multicriteria Acceptability Analysis (SMAA) and the Choquet integral preference model. As a result, the CII constructed here takes into account: (i) the hierarchical organization of the EIS; (ii) the interaction of elementary indicators; (iii) the involvement of  the Triple Helix agents preferences (University, Industry and Government) in the construction of the CII and; (iv) the consideration of robustness issues related to the stability of the results regarding the variability of the weights assigned to the dimensions of EIS.

This approach provides a more in-depth analysis of the countries innovation performance at the comprehensive level and for the specific macro-criterion incorporating the Triple Helix framework perspective. This proposal can be a valuable tool for innovation policy makers and practitioners to first, identify the strengths and weaknesses of their respective national innovation systems and, thus, second to design, deploy and develop specific policies and practices accordingly.

Finally as in our research we are assuming that a group of experts for each group  of the Triple Helix agree on the preference information that we are listing, a further direction of research could be to investigate how integrate the preferences of several experts in the decision making process. Moreover, while we are confident that the proposed methodology can also be applied to other domains (besides the measurement of innovation performance) where composite indicators are used, we acknowledge that further systematical comparisons between the EIS methodology discussed in this paper and other frameworks (such as the GII) are still needed to validate the feasibility of our approach beyond the case presented here. Further research would also benefit from testing the applicability of  the outranking approach in the aggregation stage of composite indicators by adopting  ELECTRE and PROMETHEE methods  to overcome the limitations related to the ``compensability" between criteria.

%%%%%%%%%%%%%%%%%%%%%%%%%%%%%%
%\section*{Acknowledgments}%%%
%%%%%%%%%%%%%%%%%%%%%%%%%%%%%%
%Salvatore Corrente and Salvatore Greco wish to acknowledge funding by the “FIR of the University of Catania BCAEA3, New developments in Multiple Criteria Decision Aiding (MCDA) and their application to territorial competitiveness” as well as by the research project “Data analytics for the entrepreneurial ecosystems, sustainable development and wellbeing indices”. Salvatore Greco has also benefited of the fund ``Chance'' of the university of Catania.

%\begin{acknowledgements}
%Salvatore Corrente and Salvatore Greco wish to acknowledge funding by the “FIR of the University of Catania BCAEA3, New developments in Multiple Criteria Decision Aiding (MCDA) and their application to territorial competitiveness” as well as by the research project “Data analytics for the entrepreneurial ecosystems, sustainable development and wellbeing indices”. Salvatore Greco has also benefited of the fund ``Chance'' of the university of Catania.
%\end{acknowledgements}

%%%%%%%%%%%%%%%%%%%%%%%%%%%%%
%\section*{References}%%%%%%%
%%%%%%%%%%%%%%%%%%%%%%%%%%%%%

% BibTeX users please use one of
%\bibliographystyle{spbasic}      % basic style, author-year citations
\bibliographystyle{spmpsci}      % mathematics and physical sciences
%\bibliographystyle{spphys}       % APS-like style for physics
%\bibliography{}   % name your BibTeX data base
\bibliography{innovacion}

%% Non-BibTeX users please use
%\begin{thebibliography}{}
%%
%% and use \bibitem to create references. Consult the Instructions
%% for authors for reference list style.
%%
%\bibitem{RefJ}
%% Format for Journal Reference
%Author, Article title, Journal, Volume, page numbers (year)
%% Format for books
%\bibitem{RefB}
%Author, Book title, page numbers. Publisher, place (year)
%% etc
%\end{thebibliography}
%%%%%%%%%%%%%%%%%%%%%%%%%%%%%%%%%%%%%%%%%%%%%%%%%%%%%%%%%%%%
\section{Appendix - Supplementary material}
\label{Appendix}%%
%%%%%%%%%%%%%%%%%%%%%%%%%%%%%%%%%%%%%%%%%%%%%%%%%%%%%%%%%%%%

\begin{sidewaystable}[t!]
	\centering
	\caption{Evaluations of the countries on the elementary criteria}
	\vspace{2ex}
	\resizebox{1\textheight}{!}{\input{table_performance_horizontal}}
	\label{tab:performances-countries}%
\end{sidewaystable}%

\begin{sidewaystable}[t!]
	\centering
	\caption{Normalized evaluations of the countries on the elementary criteria}
	\vspace{2ex}
	\resizebox{1\textheight}{!}{\input{table_norm_performance_horizontal}}
	\label{tab:normalized_performances-countries}%
\end{sidewaystable}%

\clearpage

%\begin{table}[h!]
%  \centering
%  \caption{Evaluations of the countries on the elementary criteria}
% \resizebox{\textwidth}{!}{\input{table_performance}}
%  \label{tab:performances-countries}%
%\end{table}%
%
%\begin{table}[h]
%  \centering
%  \caption{Normalized evaluations of the countries on the elementary criteria}
% \resizebox{\textwidth}{!}{\input{table_norm_performance}}
%  \label{tab:normalized-performances-countries}%
%\end{table}%

\end{document}

%% file: table_description_criteria.tex
{
 \begin{tabular}{rrrrl}
    \toprule
    \multicolumn{1}{l}{} &       & \multicolumn{1}{l}{} &       & \textit{Description of the elementary criteria descending from the macrocriteria HR, ARS, IFE, etc... } \\
\midrule
    \multicolumn{1}{l}{\multirow{7}[1]{*}{FC \;($g_{1}$)}} &       & \multicolumn{1}{l}{HR \;($g_{(1,1)}$)} &       & New doctorate graduates per 1000 population aged 25-34 ($g_{(1,1,1)}$)\\
          &       &       &       & Percentage population aged 25-34 having completed tertiary education \;($g_{(1,1,2)}$)\\
          \medskip
          &       &       &       & Percentage population aged 25-64 participating in lifelong learning \;($g_{(1,1,3)}$) \\ 
         
          &       & \multicolumn{1}{l}{ARS \;($g_{(1,2)}$)} &       & International scientific co-publications per million population \;($g_{(1,2,1)}$)\\
          &       &       &       & Scientific publications among the top-10\% most cited publications worldwide as percentage\\
          & & & & of total scientific publications of the country \;($g_{(1,2,2)}$)\\
          
            \medskip
          &       &       &       & Foreign doctorate students as a percentage of all doctorate students \;($g_{(1,2,3)}$) \\
          
          &       & \multicolumn{1}{l}{IFE \;($g_{(1,3)}$)} &       & Broadband penetration: Percentage of enterprises with a maximum contracted download\\
           \medskip
          & & & &speed of the fastest fixed internet connection of at least 100 Mb/s \;($g_{(1,3,1)}$)\\
     
          &       &       &       & Opportunity-driven entrepreneurship: ratio between the share of persons\\
           & & & &involved in improvement-driven entrepreneurship and the share of\\
           & & & & persons involved in necessity-driven entrepreneurship \;($g_{(1,3,2)}$) \\
\midrule
    \multicolumn{1}{l}{\multirow{4}[0]{*}{IN \;($g_{2}$)}} &       & \multicolumn{1}{l}{FS \;($g_{(2,1)}$)} &       &R\&D expenditure in the public sector (percentage of GDP) \; ($g_{(2,1,1)}$)\\
       \medskip
          &       &       &       & Venture capital investments (percentage of GDP) \; $(g_{(2,1,2)})$\\

          &       & \multicolumn{1}{l}{FI \;($g_{(2,2)}$)} &     
          & R\&D expenditure in the business sector (percentage of GDP) \; $(g_{(2,2,1)})$\\
          &       &       &       & Non-R\&D innovation expenditure (percentage of turnover) \; $(g_{(2,2,2)})$\\
       
          &       &       &       & Enterprises providing training to develop or upgrade ICT skills of their personnel \; $(g_{(2,2,3)})$\\
          \midrule
         
    \multicolumn{1}{l}{\multirow{7}[0]{*}{IA \;($g_{3}$)}} &       
    
     & \multicolumn{1}{l}{IT \; ($g_{(3,1)}$)} &       & SMEs with product or process innovations \; ($g_{(3,1,1)}$)\\
          &       &       &       & SMEs with marketing or organisational innovations \; ($g_{(3,1,2)}$)\\
           \medskip
          &       &       &       & SMEs innovating in-house \; ($g_{(3,1,3)}$)\\
          &       & \multicolumn{1}{l}{LIN \; ($g_{(3,2)}$)} &       & Innovative SMEs collaborating with others \; ($g_{(3,2,1)}$)\\
          &       &       &       & Public-private co-publications \; ($g_{(3,2,2)}$)\\
                 \medskip
          &       &       &       & Private co-funding of public R\&D expenditures \; ($g_{(3,2,3)}$)\\
    
          &       & \multicolumn{1}{l}{IAS \; ($g_{(3,3)}$)} &       &PCT patent applications \; ($g_{(3,3,1)}$)\\
          &       &       &       & Trademark applications \; ($g_{(3,3,2)}$)\\
          &       &       &       & Design applications \; ($g_{(3,3,3)}$)\\
\midrule
    \multicolumn{1}{l}{\multirow{3}[0]{*}{IMP \;$(g_{4})$}} &       & \multicolumn{1}{l}{EI\; $(g_{(4,1)})$} &       & Employment in knowledge-intensive activities \; $(g_{(4,1,1)})$\\    \medskip
          &       &       &       & Employment fast-growing firms innovative sectors \; $(g_{(4,1,2)})$ \\
              
          &       &       &       & Medium \& high-tech product exports \; $(g_{(4,2,1)})$ \\
         
          &       & \multicolumn{1}{l}{SE \; $(g_{(4,2)})$} &       & Knowledge-intensive services exports \; $(g_{(4,2,2)})$\\
          &       &       &       & Sales of new-to-market and new-to-firm innovations \; $(g_{(4,2,3)})$\\
    \hline
    \end{tabular}
    }

%% file: table_RAI_DMU.tex
% Table generated by Excel2LaTeX from sheet 'DMU'

    \begin{tabular}{llllll}
     \midrule

    (a) Comprehensive level  & Best ($b_{k,0}^{Best}$) & Worst ($b_{k,0}^{Worst}$) & $high_1$ ($b_{k,0}^{high_1}$)  & $high_2$ ($b_{k,0}^{high_2}$)  & $high_3$ ($b_{k,0}^{high_3}$)  \\
    \midrule
        Sweden (SE) & 1 (73.93\%) & 2 (26.07\%) & 1 (73.93\%) & 2 (26.07\%) & 28 (0.00\%) \\
        Denmark (DK) & 1 (26.07\%) & 2 (73.93\%) & 2 (73.93\%) & 1 (26.07\%) & 28 (0.00\%) \\
        Finland (FI) & 3 (64.61\%) & 5 (0.13\%) & 3 (64.61\%) & 4 (35.26\%) & 5 (0.13\%) \\
        Netherlands (NL) & 3 (35.39\%) & 5 (1.69\%) & 4 (62.92\%) & 3 (35.39\%) & 5 (1.69\%) \\
        Germany (DE) & 4 (0.12\%) & 10 (0.71\%) & 5 (36.17\%) & 6 (22.68\%) & 7 (19.66\%) \\
         \ldots    &   \ldots     &   \ldots     &    \ldots    &  \ldots      & \ldots  \\
        Croatia (HR) & 23 (0.37\%) & 26 (82.37\%) & 26 (82.37\%) & 25 (9.48\%) & 24 (7.79\%) \\
        Bulgaria (BG) & 27 (100.00\%) & 27 (100.00\%) & 27 (100.00\%) & 28 (0.00\%) & 26 (0.00\%) \\
        Romania (RO) & 28 (100.00\%) & 28 (100.00\%) & 28 (100.00\%) & 27 (0.00\%) & 26 (0.00\%) \\
    \midrule
   (b) Framework Conditions (FC) & Best ($b_{k,1}^{Best}$) & Worst ($b_{k,1}^{Worst}$) & $high_1$ ($b_{k,1}^{high_1}$)  & $high_2$ ($b_{k,1}^{high_2}$)  & $high_3$ ($b_{k,1}^{high_3}$)  \\
     \midrule
  
        Sweden (SE) & 1 (0.01\%) & 2 (99.99\%) & 2 (99.99\%) & 1 (0.01\%) & 28 (0.00\%) \\							
        Denmark (DK) & 1 (99.99\%) & 2 (0.01\%) & 1 (99.99\%) & 2 (0.01\%) & 28 (0.00\%) \\							
        Finland (FI) & 3 (36.70\%) & 7 (2.92\%) & 3 (36.70\%) & 4 (24.17\%) & 5 (22.71\%) \\							
        Netherlands (NL) & 3 (12.13\%) & 6 (0.00\%) & 4 (48.60\%) & 5 (39.27\%) & 3 (12.13\%) \\							
        Germany (DE) & 12 (10.19\%) & 18 (0.59\%) & 16 (32.91\%) & 15 (23.30\%) & 14 (19.07\%) \\							
     \ldots    &   \ldots     &   \ldots     &    \ldots    &  \ldots      & \ldots  \\
        Croatia (HR) & 25 (17.43\%) & 28 (4.89\%) & 26 (71.16\%) & 25 (17.43\%) & 27 (6.51\%) \\							
        Bulgaria (BG) & 26 (6.37\%) & 28 (15.75\%) & 27 (77.88\%) & 28 (15.75\%) & 26 (6.37\%) \\							
        Romania (RO) & 25 (0.23\%) & 28 (78.97\%) & 28 (78.97\%) & 27 (14.51\%) & 26 (6.29\%) \\							
    \midrule
    (c) Investments (IN) & Best ($b_{k,2}^{Best}$) & Worst ($b_{k,2}^{Worst}$) & $high_1$ ($b_{k,2}^{high_1}$)  & $high_2$ ($b_{k,2}^{high_2}$)  & $high_3$ ($b_{k,2}^{high_3}$) \\
    
    \midrule
    
    Sweden (SE) & 1 (10.81\%) & 5 (2.03\%) & 2 (48.00\%) & 3 (22.44\%) & 4 (16.73\%) \\
    Denmark (DK) & 1 (80.26\%) & 4 (0.25\%) & 1 (80.26\%) & 2 (16.51\%) & 3 (2.99\%) \\
    Finland (FI) & 1 (8.92\%) & 5 (0.13\%) & 3 (43.75\%) & 2 (31.91\%) & 4 (15.29\%) \\
    Netherlands (NL) & 5 (3.15\%) & 10 (0.07\%) & 6 (57.84\%) & 7 (25.69\%) & 8 (10.76\%) \\
    Germany (DE) & 1 (0.01\%) & 5 (36.30\%) & 4 (48.00\%) & 5 (36.30\%) & 3 (12.58\%) \\
     \ldots    &   \ldots     &   \ldots     &    \ldots    &  \ldots      & \ldots  \\
    Croatia (HR) & 16 (0.25\%) & 23 (3.41\%) & 19 (29.25\%) & 20 (21.81\%) & 21 (19.70\%) \\
    Bulgaria (BG) & 26 (0.20\%) & 27 (99.80\%) & 27 (99.80\%) & 26 (0.20\%) & 28 (0.00\%) \\
    Romania (RO) & 28 (100.00\%) & 28 (100.00\%) & 28 (100.00\%) & 27 (0.00\%) & 26 (0.00\%) \\
    \midrule
    (d) Innovation Activities (IA) & Best ($b_{k,3}^{Best}$) & Worst ($b_{k,3}^{Worst}$) & $high_1$ ($b_{k,3}^{high_1}$)  & $high_2$ ($b_{k,3}^{high_2}$)  & $high_3$ ($b_{k,3}^{high_3}$)\\
    
    \midrule
   
    Sweden (SE) & 1 (2.76\%) & 9 (0.21\%) & 6 (37.67\%) & 5 (19.75\%) & 7 (12.85\%) \\
    Denmark (DK) & 1 (2.86\%) & 9 (5.31\%) & 7 (42.14\%) & 8 (18.56\%) & 6 (12.72\%) \\
    Finland (FI) & 2 (6.73\%) & 7 (0.39\%) & 4 (35.35\%) & 5 (28.39\%) & 3 (22.27\%) \\
    Netherlands (NL) & 1 (0.57\%) & 8 (0.20\%) & 3 (37.36\%) & 4 (22.97\%) & 5 (18.04\%) \\
    Germany (DE) & 1 (23.84\%) & 9 (0.92\%) & 1 (23.84\%) & 2 (18.78\%) & 5 (13.07\%) \\

     \ldots    &   \ldots     &   \ldots     &    \ldots    &  \ldots      & \ldots  \\

    Croatia (HR) & 20 (4.85\%) & 23 (0.52\%) & 22 (62.20\%) & 21 (32.43\%) & 20 (4.85\%) \\					
    Bulgaria (BG) & 22 (0.32\%) & 26 (30.91\%) & 25 (47.49\%) & 26 (30.91\%) & 24 (13.14\%) \\					
    Romania (RO) & 28 (100.00\%) & 28 (100.00\%) & 28 (100.00\%) & 27 (0.00\%) & 26 (0.00\%) \\

    \midrule
    (e) Impacts (IMP) &    Best ($b_{k,4}^{Best}$) & Worst ($b_{k,4}^{Worst}$) & $high_1$ ($b_{k,4}^{high_1}$)  & $high_2$ ($b_{k,4}^{high_2}$)  & $high_3$ ($b_{k,4}^{high_3}$) \\
    \midrule
       Sweden (SE) & 3 (1.01\%) & 9 (0.02\%) & 4 (58.67\%) & 5 (22.78\%) & 6 (14.29\%) \\							
       Denmark (DK) & 7 (0.70\%) & 15 (0.31\%) & 12 (45.04\%) & 11 (14.91\%) & 9 (12.46\%) \\							
       Finland (FI) & 10 (4.11\%) & 21 (0.24\%) & 13 (20.69\%) & 14 (18.92\%) & 15 (10.01\%) \\							
       Netherlands (NL) & 4 (4.99\%) & 9 (5.20\%) & 5 (40.53\%) & 6 (20.22\%) & 7 (14.79\%) \\							
       Germany (DE) & 4 (0.45\%) & 9 (10.35\%) & 6 (31.79\%) & 7 (31.01\%) & 8 (18.63\%) \\							
        \ldots    &   \ldots     &   \ldots     &    \ldots    &  \ldots      & \ldots  \\
       Croatia (HR) & 25 (1.50\%) & 28 (39.30\%) & 28 (39.30\%) & 27 (31.56\%) & 26 (27.64\%) \\							
       Bulgaria (BG) & 12 (0.00\%) & 27 (0.15\%) & 24 (25.01\%) & 14 (17.80\%) & 25 (16.79\%) \\							
       Romania (RO) & 22 (0.05\%) & 28 (40.02\%) & 28 (40.02\%) & 26 (34.47\%) & 27 (15.27\%) \\							
    \midrule
    \end{tabular}%

%% file: table_RAI_DMI.tex
    \begin{tabular}{llllll}
     \midrule

    (a) Comprehensive level  & Best ($b_{k,0}^{Best}$) & Worst ($b_{k,0}^{Worst}$) & $high_1$ ($b_{k,0}^{high_1}$)  & $high_2$ ($b_{k,0}^{high_2}$)  & $high_3$ ($b_{k,0}^{high_3}$)  \\
	 \midrule
	 
    Sweden (SE) & 1 (100.00\%) & 1 (100.00\%) & 1 (100.00\%) & 28 (0.00\%) & 27 (0.00\%) \\
    Finland (FI) & 2 (59.50\%) & 5 (0.56\%) & 2 (59.50\%) & 3 (36.33\%) & 4 (3.61\%) \\
    Netherlands (NL) & 2 (1.02\%) & 7 (4.47\%) & 4 (53.21\%) & 5 (26.09\%) & 3 (10.12\%) \\
    United Kingdom (UK) & 2 (11.85\%) & 8 (0.32\%) & 5 (22.07\%) & 6 (21.82\%) & 3 (17.52\%) \\
    Denmark (DK) & 2 (8.74\%) & 7 (13.02\%) & 5 (27.38\%) & 3 (21.84\%) & 6 (17.69\%) \\
     \ldots    &   \ldots     &   \ldots     &    \ldots    &  \ldots      & \ldots  \\
    Poland (PL) & 25 (31.15\%) & 26 (68.85\%) & 26 (68.85\%) & 25 (31.15\%) & 28 (0.00\%) \\
    Bulgaria (BG) & 27 (98.94\%) & 28 (1.06\%) & 27 (98.94\%) & 28 (1.06\%) & 26 (0.00\%) \\
    Romania (RO) & 27 (1.06\%) & 28 (98.94\%) & 28 (98.94\%) & 27 (1.06\%) & 26 (0.00\%) \\
    \midrule

    (b) Framework Conditions (FC) & Best ($b_{k,1}^{Best}$) & Worst ($b_{k,1}^{Worst}$) & $high_1$ ($b_{k,1}^{high_1}$)  & $high_2$ ($b_{k,1}^{high_2}$)  & $high_3$ ($b_{k,1}^{high_3}$)  \\

    \midrule
    Sweden (SE) & 2 (93.76\%) & 3 (6.24\%) & 2 (93.76\%) & 3 (6.24\%) & 28 (0.00\%) \\
    Finland (FI) & 3 (22.12\%) & 7 (0.56\%) & 4 (38.03\%) & 5 (33.71\%) & 3 (22.12\%) \\
    Netherlands (NL) & 3 (7.33\%) & 6 (0.16\%) & 4 (50.80\%) & 5 (41.71\%) & 3 (7.33\%) \\
    United Kingdom (UK) & 4 (0.16\%) & 8 (0.42\%) & 6 (69.30\%) & 7 (24.99\%) & 5 (5.12\%) \\
    Denmark (DK) & 1 (98.29\%) & 2 (1.71\%) & 1 (98.29\%) & 2 (1.71\%) & 28 (0.00\%) \\
     \ldots    &   \ldots     &   \ldots     &    \ldots    &  \ldots      & \ldots  \\
    Poland (PL) & 21 (0.80\%) & 27 (0.33\%) & 25 (90.92\%) & 26 (4.11\%) & 24 (2.96\%) \\
    Bulgaria (BG) & 25 (0.75\%) & 28 (40.08\%) & 28 (40.08\%) & 27 (33.06\%) & 26 (26.11\%) \\
    Romania (RO) & 25 (0.82\%) & 28 (33.98\%) & 26 (38.17\%) & 28 (33.98\%) & 27 (27.03\%) \\
    \midrule

    (c) Investments (IN) & Best ($b_{k,2}^{Best}$) & Worst ($b_{k,2}^{Worst}$) & $high_1$ ($b_{k,2}^{high_1}$)  & $high_2$ ($b_{k,2}^{high_2}$)  & $high_3$ ($b_{k,2}^{high_3}$) \\
    
    \midrule
    
    Sweden (SE) & 1 (12.89\%) & 5 (0.64\%) & 2 (46.28\%) & 3 (27.77\%) & 1 (12.89\%) \\
    Finland (FI) & 1 (59.91\%) & 5 (0.07\%) & 1 (59.91\%) & 2 (26.76\%) & 4 (7.07\%) \\
    Netherlands (NL) & 4 (0.62\%) & 12 (0.17\%) & 9 (26.05\%) & 10 (18.77\%) & 5 (17.59\%) \\
    United Kingdom (UK) & 3 (12.75\%) & 10 (0.04\%) & 4 (43.47\%) & 5 (21.65\%) & 6 (13.92\%) \\
    Denmark (DK) & 3 (0.17\%) & 15 (5.52\%) & 8 (18.82\%) & 12 (14.76\%) & 10 (13.55\%) \\
     \ldots    &   \ldots     &   \ldots     &    \ldots    &  \ldots      & \ldots  \\

    Poland (PL) & 16 (0.50\%) & 23 (3.23\%) & 22 (55.99\%) & 20 (10.79\%) & 18 (9.56\%) \\
    Bulgaria (BG) & 24 (18.06\%) & 27 (12.92\%) & 25 (47.41\%) & 26 (21.61\%) & 24 (18.06\%) \\
    Romania (RO) & 27 (1.67\%) & 28 (98.33\%) & 28 (98.33\%) & 27 (1.67\%) & 26 (0.00\%) \\
    \midrule

  (d) Innovation Activities (IA) & Best ($b_{k,3}^{Best}$) & Worst ($b_{k,3}^{Worst}$) & $high_1$ ($b_{k,3}^{high_1}$)  & $high_2$ ($b_{k,3}^{high_2}$)  & $high_3$ ($b_{k,3}^{high_3}$)\\

    \midrule
    Sweden (SE) & 1 (0.37\%) & 9 (0.01\%) & 6 (45.63\%) & 5 (31.10\%) & 4 (9.58\%) \\
    Finland (FI) & 2 (12.52\%) & 7 (0.88\%) & 4 (47.23\%) & 3 (20.25\%) & 5 (16.57\%) \\
    Netherlands (NL) & 2 (2.78\%) & 8 (0.06\%) & 3 (52.34\%) & 4 (19.57\%) & 5 (11.84\%) \\
    United Kingdom (UK) & 5 (0.04\%) & 15 (0.74\%) & 8 (35.68\%) & 9 (26.49\%) & 10 (11.65\%) \\
    Denmark (DK) & 1 (3.20\%) & 10 (1.47\%) & 7 (62.57\%) & 6 (10.63\%) & 8 (8.73\%) \\
     \ldots    &   \ldots     &   \ldots     &    \ldots    &  \ldots      & \ldots  \\

    Poland (PL) & 26 (16.33\%) & 27 (83.67\%) & 27 (83.67\%) & 26 (16.33\%) & 28 (0.00\%) \\
    Bulgaria (BG) & 24 (1.04\%) & 26 (55.27\%) & 26 (55.27\%) & 25 (43.69\%) & 24 (1.04\%) \\
    Romania (RO) & 28 (100.00\%) & 28 (100.00\%) & 28 (100.00\%) & 27 (0.00\%) & 26 (0.00\%) \\
    \midrule

   (e) Impacts (IMP) &    Best ($b_{k,4}^{Best}$) & Worst ($b_{k,4}^{Worst}$) & $high_1$ ($b_{k,4}^{high_1}$)  & $high_2$ ($b_{k,4}^{high_2}$)  & $high_3$ ($b_{k,4}^{high_3}$) \\

    \midrule
    Sweden (SE) & 4 (17.80\%) & 9 (11.71\%) & 5 (31.69\%) & 6 (18.39\%) & 4 (17.80\%) \\
    Finland (FI) & 10 (0.00\%) & 20 (0.60\%) & 17 (24.52\%) & 15 (22.34\%) & 18 (21.07\%) \\
    Netherlands (NL) & 5 (1.98\%) & 13 (1.29\%) & 6 (26.26\%) & 11 (23.41\%) & 7 (13.11\%) \\
    United Kingdom (UK) & 1 (1.93\%) & 6 (0.00\%) & 2 (52.43\%) & 3 (36.50\%) & 4 (5.75\%) \\
    Denmark (DK) & 7 (0.06\%) & 16 (2.27\%) & 13 (24.64\%) & 14 (20.41\%) & 12 (18.12\%) \\
     \ldots    &   \ldots     &   \ldots     &    \ldots    &  \ldots      & \ldots  \\
    Poland (PL) & 14 (0.04\%) & 23 (0.44\%) & 21 (46.06\%) & 19 (23.53\%) & 20 (21.05\%) \\
    Bulgaria (BG) & 21 (0.29\%) & 28 (5.94\%) & 27 (34.62\%) & 26 (29.00\%) & 25 (20.58\%) \\
    Romania (RO) & 18 (0.01\%) & 27 (0.58\%) & 22 (49.53\%) & 23 (14.05\%) & 25 (8.22\%) \\
    
    \midrule
    
    \end{tabular}%

%% file: table_RAI_DMG.tex
    \begin{tabular}{llllll}
    \midrule

    (a) Comprehensive level  & Best ($b_{k,0}^{Best}$) & Worst ($b_{k,0}^{Worst}$) & $high_1$ ($b_{k,0}^{high_1}$)  & $high_2$ ($b_{k,0}^{high_2}$)  & $high_3$ ($b_{k,0}^{high_3}$)  \\
   	
    \midrule
    Sweden (SE) & 1 (71.99\%) & 3 (1.44\%) & 1 (71.99\%) & 2 (26.58\%) & 3 (1.44\%) \\
    United Kingdom (UK) & 1 (28.01\%) & 6 (0.34\%) & 2 (48.51\%) & 1 (28.01\%) & 4 (10.52\%) \\
    Denmark (DK) & 2 (22.77\%) & 6 (0.29\%) & 3 (29.89\%) & 5 (29.08\%) & 2 (22.77\%) \\
    Netherlands (NL) & 2 (0.71\%) & 6 (3.55\%) & 4 (45.68\%) & 3 (29.81\%) & 5 (20.26\%) \\
    Ireland (IE) & 2 (1.44\%) & 8 (0.71\%) & 3 (27.34\%) & 5 (21.85\%) & 4 (20.74\%) \\
    \ldots    &   \ldots     &   \ldots     &    \ldots    &  \ldots      & \ldots  \\
    Croatia (HR) & 25 (22.04\%) & 26 (77.96\%) & 26 (77.96\%) & 25 (22.04\%) & 28 (0.00\%) \\
    Bulgaria (BG) & 27 (100.00\%) & 27 (100.00\%) & 27 (100.00\%) & 28 (0.00\%) & 26 (0.00\%) \\
    Romania (RO) & 28 (100.00\%) & 28 (100.00\%) & 28 (100.00\%) & 27 (0.00\%) & 26 (0.00\%) \\
    \midrule
    
   (b) Framework Conditions (FC) & Best ($b_{k,1}^{Best}$) & Worst ($b_{k,1}^{Worst}$) & $high_1$ ($b_{k,1}^{high_1}$)  & $high_2$ ($b_{k,1}^{high_2}$)  & $high_3$ ($b_{k,1}^{high_3}$)  \\

    \midrule
    Sweden (SE) & 2 (86.00\%) & 3 (14.00\%) & 2 (86.00\%) & 3 (14.00\%) & 28 (0.00\%) \\
    United Kingdom (UK) & 4 (1.94\%) & 7 (11.19\%) & 6 (71.88\%) & 5 (15.00\%) & 7 (11.19\%) \\
    Denmark (DK) & 1 (96.03\%) & 2 (3.97\%) & 1 (96.03\%) & 2 (3.97\%) & 28 (0.00\%) \\
    Netherlands (NL) & 3 (0.13\%) & 6 (0.19\%) & 5 (56.09\%) & 4 (43.59\%) & 6 (0.19\%) \\
    Ireland (IE) & 7 (0.56\%) & 11 (0.54\%) & 10 (46.75\%) & 9 (38.22\%) & 8 (13.93\%) \\
   \ldots    &   \ldots     &   \ldots     &    \ldots    &  \ldots      & \ldots  \\
    Croatia (HR) & 25 (3.13\%) & 28 (26.34\%) & 26 (51.66\%) & 28 (26.34\%) & 27 (18.86\%) \\
    Bulgaria (BG) & 25 (0.01\%) & 28 (20.42\%) & 27 (59.88\%) & 28 (20.42\%) & 26 (19.68\%) \\
    Romania (RO) & 25 (1.26\%) & 28 (52.84\%) & 28 (52.84\%) & 26 (25.10\%) & 27 (20.80\%) \\
    \midrule
    
     (c) Investments (IN) & Best ($b_{k,2}^{Best}$) & Worst ($b_{k,2}^{Worst}$) & $high_1$ ($b_{k,2}^{high_1}$)  & $high_2$ ($b_{k,2}^{high_2}$)  & $high_3$ ($b_{k,2}^{high_3}$) \\
    
    \midrule
    
    Sweden (SE) & 1 (5.45\%) & 5 (0.85\%) & 3 (47.25\%) & 2 (31.45\%) & 4 (15.01\%) \\
    United Kingdom (UK) & 2 (4.47\%) & 10 (0.54\%) & 4 (54.20\%) & 3 (15.36\%) & 5 (11.42\%) \\
    Denmark (DK) & 3 (0.35\%) & 15 (0.57\%) & 12 (15.82\%) & 10 (15.80\%) & 8 (15.56\%) \\
    Netherlands (NL) & 3 (0.11\%) & 12 (0.13\%) & 5 (35.37\%) & 9 (19.51\%) & 8 (14.68\%) \\
    Ireland (IE) & 6 (1.13\%) & 16 (0.01\%) & 10 (23.77\%) & 11 (17.28\%) & 12 (17.11\%) \\
    \ldots    &   \ldots     &   \ldots     &    \ldots    &  \ldots      & \ldots  \\
    Croatia (HR) & 7 (0.11\%) & 20 (0.03\%) & 15 (52.91\%) & 16 (34.59\%) & 17 (4.15\%) \\
    Bulgaria (BG) & 24 (1.37\%) & 27 (25.74\%) & 25 (49.44\%) & 27 (25.74\%) & 26 (23.45\%) \\
    Romania (RO) & 27 (3.16\%) & 28 (96.84\%) & 28 (96.84\%) & 27 (3.16\%) & 26 (0.00\%) \\
    \midrule
    
   (d) Innovation Activities (IA) & Best ($b_{k,3}^{Best}$) & Worst ($b_{k,3}^{Worst}$) & $high_1$ ($b_{k,3}^{high_1}$)  & $high_2$ ($b_{k,3}^{high_2}$)  & $high_3$ ($b_{k,3}^{high_3}$)\\

    \midrule
    Sweden (SE) & 1 (3.39\%) & 8 (1.95\%) & 6 (42.44\%) & 7 (19.02\%) & 5 (17.85\%) \\
    United Kingdom (UK) & 5 (0.38\%) & 15 (0.30\%) & 9 (35.67\%) & 8 (18.60\%) & 10 (18.19\%) \\
    Denmark (DK) & 1 (4.32\%) & 11 (0.06\%) & 7 (41.66\%) & 8 (17.47\%) & 6 (9.59\%) \\
    Netherlands (NL) & 2 (6.37\%) & 8 (0.52\%) & 3 (41.54\%) & 4 (20.72\%) & 5 (15.22\%) \\
    Ireland (IE) & 9 (1.86\%) & 16 (3.94\%) & 11 (25.36\%) & 12 (18.96\%) & 10 (14.96\%) \\
    \ldots    &   \ldots     &   \ldots     &    \ldots    &  \ldots      & \ldots  \\
    Croatia (HR) & 21 (1.31\%) & 23 (3.59\%) & 22 (95.10\%) & 23 (3.59\%) & 21 (1.31\%) \\
    Bulgaria (BG) & 23 (2.37\%) & 26 (44.19\%) & 26 (44.19\%) & 25 (38.63\%) & 24 (14.80\%) \\
    Romania (RO) & 28 (100.00\%) & 28 (100.00\%) & 28 (100.00\%) & 27 (0.00\%) & 26 (0.00\%) \\
    \midrule
    
    (e) Impacts (IMP) &    Best ($b_{k,4}^{Best}$) & Worst ($b_{k,4}^{Worst}$) & $high_1$ ($b_{k,4}^{high_1}$)  & $high_2$ ($b_{k,4}^{high_2}$)  & $high_3$ ($b_{k,4}^{high_3}$) \\

    \midrule
    Sweden (SE) & 4 (0.80\%) & 14 (0.36\%) & 9 (37.61\%) & 10 (18.47\%) & 6 (14.26\%) \\
    United Kingdom (UK) & 1 (13.67\%) & 3 (2.80\%) & 2 (83.53\%) & 1 (13.67\%) & 3 (2.80\%) \\
    Denmark (DK) & 7 (4.70\%) & 19 (2.78\%) & 15 (20.86\%) & 16 (15.60\%) & 17 (12.43\%) \\
    Netherlands (NL) & 4 (12.56\%) & 11 (0.21\%) & 8 (51.78\%) & 5 (14.36\%) & 4 (12.56\%) \\
    Ireland (IE) & 1 (86.33\%) & 2 (13.67\%) & 1 (86.33\%) & 2 (13.67\%) & 28 (0.00\%) \\
     \ldots    &   \ldots     &   \ldots     &    \ldots    &  \ldots      & \ldots  \\
    Croatia (HR) & 27 (28.98\%) & 28 (71.02\%) & 28 (71.02\%) & 27 (28.98\%) & 26 (0.00\%) \\
    Bulgaria (BG) & 17 (0.14\%) & 27 (2.00\%) & 22 (18.71\%) & 25 (17.70\%) & 26 (16.59\%) \\
    Romania (RO) & 20 (0.01\%) & 27 (9.18\%) & 26 (53.40\%) & 25 (13.72\%) & 27 (9.18\%) \\
    \bottomrule
    \end{tabular}%
 

%% file: table_summary.tex
% Table generated by Excel2LaTeX from sheet 'Hoja1'
{
    \begin{tabular}{lcrrc}
    \toprule
    Country & University & \multicolumn{1}{c}{Industry} & \multicolumn{1}{c}{Government} & EIS \\
    \midrule
    Belgium (BE) & 9     & {7} & 10    & 9 \\
    Bulgaria (BG) & 27    & 27    & 27    & 27 \\
    Czech Republic (CZ) & 13    & 14    & 14    & 13 \\
    Denmark (DK) & 2     & {5} & 3     & 2 \\
    Germany (DE) & 5     & 6     & {9} & 6 \\
    Estonia (EE) & {14}    & 12    & 12    & 15 \\
    Ireland (IE) & 11    & 10    & {5} & 10 \\
    Grece (EL) & {20}    & 25    & 24    & 22 \\
    Spain (ES) & 17    & {19} & 15    & 17 \\
    France (FR) & 10    & 11    & {8} & 11 \\
    Croatia (HR) & 26    & 24    & 26    & 26 \\
    Italy (IT) & 22    & 21    & 22    & 19 \\
    Cyprus (CY) & {21} & 17    & 18    & 20 \\
    Latvia (LV) & 24    & 23    & 23    & 24 \\
    Lithuania (LT) & {16} & 18    & 19    & 16 \\
    Luxembourg (LU) & 8     & 8     & 7     & 8 \\
    Hungary (HU) & {23} & 16    & 16    & 23 \\
    Malta (MT) & 19    & 20    & 21    & 18 \\
    Netherland (NL) & 4     & 3     & 4     & 4 \\
    Austria (AT) & {6} & 9     & 11    & 7 \\
    Poland (PL) & 25    & 26    & 25    & 25 \\
    Portugal (PT) & 15    & 15    & 17    & 14 \\
    Romania (RO) & 28    & 28    & 28    & 28 \\
    Slovenia (SI) & 12    & 13    & 13    & 12 \\
    Slovakia (SK) & {18} & 22    & 20    & 21 \\
    Finland (FI) & 3     & 2     & {6} & 3 \\
    Sweden (SE) & 1     & 1     & 1     & 1 \\
    United Kingdom (UK) & {7} & 4     & 2     & 5 \\
    \bottomrule
    \end{tabular}%
 
}

%% file: table_discrepancies.tex
% Table generated by Excel2LaTeX from sheet 'Hoja1'

\begin{tabular}{lccc}
    \toprule
    Country & University & Industry & Government \\
    \midrule
    Belgium (BE) &       &     &  $\downarrow$\\
    Denmark (DK) &       & $\downarrow$     &  \\
    Germany (DE) &       &       & $\downarrow$  \\
    Estonia (EE) &    $\uparrow$    &  $\uparrow$      &  \\
    Ireland (IE) &       &       & $\uparrow$ \\
    Greece (EL) &     &    $\downarrow$    &  \\
    Spain (ES) &       & $\downarrow$      & $\uparrow$ \\
    France (FR) &       &       & $\uparrow$  \\
    Cyprus (CY) &     &   $\uparrow$    &  \\
    Lithuania (LT) &       &       & $\downarrow$  \\
    Hungary (HU) &      &    $\uparrow$   & $\uparrow$ \\
    Malta (MT) &      &       & $\downarrow$ \\
    Austria (AT) &       &       &  $\downarrow$\\
    Slovakia (SK) & $\uparrow$      &       &  \\
    Finland (FI) &       &       & $\downarrow$ \\
    United Kingdom (UK) &       &       &  $\uparrow$ \\
    \bottomrule
\end{tabular}%
 

%% file: table_performance_horizontal.tex
    \begin{tabular}{lccccccccccccccccccccccccccc}
    \toprule
          & \multicolumn{8}{c}{FC \;$\left(g_{1}\right)$} & \multicolumn{5}{c}{IN \;$\left(g_{2}\right)$} & \multicolumn{9}{c}{IA $\left(g_{(3)}\right)$} & \multicolumn{5}{c}{IMP $\left(g_{(4)}\right)$} \\
    \midrule
          & \multicolumn{3}{c}{HR \;$\left(g_{(1,1)}\right)$} & \multicolumn{3}{c}{ARS \;$\left(g_{(1,2)}\right)$} & \multicolumn{2}{c}{IFE \;$\left(g_{(1,3)}\right)$} & \multicolumn{3}{c}{FS \;$\left(g_{(2,1)}\right)$} & \multicolumn{2}{c}{FI \;$\left(g_{(2,2)}\right)$} & \multicolumn{3}{c}{FS \;$\left(g_{(3,1)}\right)$} & \multicolumn{3}{c}{LIN \;$\left(g_{(3,2)}\right)$} & \multicolumn{3}{c}{IAS \;$\left(g_{(3,3)}\right)$} & \multicolumn{2}{c}{EI \;$\left(g_{(4,1)}\right)$} & \multicolumn{3}{c}{FS \;$\left(g_{(4,2)}\right)$} \\
    \midrule
          & $g_{(1,1,1)}$   & $g_{(1,1,2)}$   & $g_{(1,1,3)}$   & $g_{(1,2,1)}$  & $g_{(1,2,2)}$  & $g_{(1,2,3)}$  & $g_{(1,3,1)}$   & $g_{(1,3,2)}$   & $g_{(2,1,1)}$   & $g_{(2,1,2)}$   & $g_{(2,2,1)}$   & $g_{(2,2,2)}$   & $g_{(2,2,3)}$   & $g_{(3,1,1)}$   & $g_{(3,1,2)}$   & $g_{(3,1,3)}$   & $g_{(3,2,1)}$ & $g_{(3,2,2)}$  & $g_{(3,2,3)}$ & $g_{(3,3,1)}$ & $g_{(3,3,2)}$  &$g_{(3,3,3)}$  & $g_{(4,1,1)}$  & $g_{(4,1,2)}$   & $g_{(4,2,1)}$   & $g_{(4,2,2)}$  & \multicolumn{1}{l}{$g_{(4,2,3)}$} \\
     \midrule      
    Belgium (BE) & 1.79  & 44.30 & 7.00  & 1408.08 & 12.63 & 42.31 & 23.00 & 1.51  & 0.68  & 0.072 & 1.77  & 0.56  & 34    & 48.26 & 45.14 & 39.75 & 28.59 & 61.05 & 0.07  & 3.35  & 7.75  & 2.75  & 15.20 & 2.46  & 48.53 & 67.86 & 7.60 \\
    Bulgaria (BG) & 1.48  & 32.80 & 2.20  & 202.41 & 4.04  & 5.14  & 10.00 & 0.99  & 0.25  & 0.015 & 0.7   & 0.74  & 8     & 14.04 & 14.75 & 11.19 & 3.11  & 1.11  & 0.01  & 0.65  & 9.49  & 7.02  & 10.40 & 6.14  & 31.04 & 41.38 & 4.80 \\
    Czech Republic (CZ) & 1.68  & 32.60 & 8.80  & 688.47 & 7.05  & 14.76 & 10.00 & 2.65  & 0.88  & 0.013 & 1.06  & 0.94  & 22    & 30.83 & 25.74 & 27.98 & 10.03 & 10.25 & 0.03  & 1.08  & 5.14  & 2.62  & 12.80 & 4.95  & 64.08 & 41.96 & 14.57 \\
    Denmark (DK) & 3.24  & 45.30 & 27.70 & 2228.92 & 13.44 & 32.06 & 31.00 & 11.09 & 1.15  & 0.059 & 1.87  & 0.29  & 28    & 34.65 & 39.98 & 28.22 & 13.23 & 131.99 & 0.02  & 6.14  & 11.60 & 7.93  & 15.80 & 4.31  & 47.84 & 74.79 & 6.96 \\
    Germany (DE) & 2.85  & 30.50 & 8.50  & 778.17 & 11.40 & 9.12  & 12.00 & 2.92  & 0.93  & 0.049 & 1.95  & 1.26  & 29    & 41.56 & 49.09 & 37.90 & 10.10 & 45.30 & 0.12  & 6.35  & 9.34  & 6.18  & 14.80 & 4.52  & 67.65 & 74.69 & 13.34 \\
    Estonia (EE) & 1.08  & 41.20 & 15.70 & 1029.68 & 8.04  & 8.30  & 12.00 & 3.40  & 0.78  & 0.136 & 0.69  & 0.85  & 13    & 17.36 & 15.03 & 15.80 & 10.76 & 1.52  & 0.04  & 1.36  & 14.97 & 3.74  & 12.70 & 3.03  & 42.72 & 45.35 & 10.49 \\
    Ireland (IE) & 2.51  & 51.80 & 6.40  & 1196.68 & 12.14 & 23.10 & 15.00 & 2.28  & 0.33  & 0.086 & 1.09  & 0.47  & 30    & 45.72 & 52.52 & 41.33 & 13.95 & 23.55 & 0.01  & 2.49  & 5.40  & 1.02  & 19.80 & 8.76  & 52.52 & 93.98 & 18.07 \\
    Grece (EL) & 1.13  & 41.00 & 4.00  & 590.80 & 8.90  & 19.85 & 2.00  & 1.17  & 0.63  & 0.001 & 0.32  & 0.76  & 15    & 34.61 & 40.14 & 31.40 & 14.76 & 7.92  & 0.04  & 0.55  & 4.58  & 1.30  & 12.20 & 4.55  & 22.69 & 44.44 & 12.75 \\
    Spain (ES) & 1.91  & 41.00 & 9.40  & 701.40 & 9.67  & 11.92 & 20.00 & 1.61  & 0.57  & 0.043 & 0.64  & 0.36  & 23    & 18.60 & 25.52 & 14.47 & 6.68  & 11.37 & 0.03  & 1.57  & 9.13  & 3.08  & 12.30 & 3.52  & 47.79 & 43.30 & 15.94 \\
    France (FR) & 1.70  & 44.00 & 18.80 & 700.18 & 11.24 & 40.05 & 9.00  & 5.31  & 0.74  & 0.083 & 1.45  & 0.50  & 20    & 35.47 & 41.62 & 31.55 & 13.21 & 32.20 & 0.04  & 4.17  & 5.85  & 2.88  & 14.20 & 4.26  & 58.61 & 67.05 & 15.02 \\
    Croatia (HR) & 1.57  & 33.00 & 3.00  & 465.56 & 4.55  & 3.16  & 6.00  & 0.98  & 0.42  & 0.054 & 0.44  & 1.20  & 22    & 25.43 & 30.84 & 21.13 & 6.78  & 5.68  & 0.03  & 0.63  & 4.22  & 0.86  & 11.70 & 2.84  & 37.98 & 19.05 & 4.91 \\
    Italy (IT) & 1.53  & 25.60 & 8.30  & 596.45 & 10.19 & 13.16 & 5.00  & 2.72  & 0.56  & 0.022 & 0.74  & 0.57  & 12    & 32.67 & 34.60 & 30.52 & 6.72  & 15.20 & 0.01  & 2.17  & 8.14  & 6.41  & 13.90 & 2.65  & 52.07 & 50.36 & 10.06 \\
    Cyprus (CY) & 0.55  & 56.30 & 6.90  & 1139.90 & 10.23 & 11.36 & 3.00  & 2.00  & 0.3   & 0.071 & 0.08  & 0.21  & 22    & 32.84 & 31.11 & 30.48 & 11.67 & 7.08  & 0.00  & 0.80  & 41.39 & 3.34  & 16.30 & 0.77  & 43.18 & 68.40 & 4.49 \\
    Latvia (LV) & 0.91  & 42.10 & 7.30  & 264.10 & 4.14  & 8.82  & 22.00 & 3.51  & 0.47  & 0.098 & 0.15  & 0.58  & 12    & 11.89 & 18.97 & 10.18 & 2.78  & 0.49  & 0.05  & 0.31  & 7.01  & 1.75  & 11.10 & 4.82  & 34.66 & 52.98 & 5.31 \\
    Lithuania (LT) & 1.12  & 54.90 & 6.00  & 392.93 & 3.96  & 3.91  & 21.00 & 2.23  & 0.76  & 0.082 & 0.28  & 2.01  & 10    & 33.69 & 24.00 & 30.38 & 15.19 & 0.68  & 0.09  & 0.80  & 6.24  & 1.42  & 9.70  & 4.02  & 34.45 & 21.00 & 8.57 \\
    Luxembourg (LU) & 1.01  & 51.50 & 16.80 & 1714.54 & 11.61 & 86.99 & 21.00 & 5.15  & 0.64  & 0.047 & 0.67  & 0.13  & 29    & 36.95 & 54.35 & 32.24 & 9.18  & 8.88  & 0.01  & 1.91  & 38.51 & 12.40 & 22.70 & 4.24  & 52.17 & 91.18 & 6.54 \\
    Hungary (HU) & 0.96  & 30.40 & 6.30  & 445.25 & 6.23  & 7.17  & 12.00 & 1.96  & 0.35  & 0.055 & 1.01  & 0.75  & 16    & 15.07 & 15.22 & 11.74 & 6.19  & 23.24 & 0.03  & 1.32  & 3.91  & 0.93  & 12.20 & 7.60  & 69.62 & 47.27 & 12.47 \\
    Malta (MT) & 0.48  & 34.00 & 7.50  & 554.78 & 9.48  & 12.39 & 12.00 & 3.14  & 0.39  & 0.000 & 0.37  & 0.36  & 23    & 26.71 & 30.78 & 23.87 & 4.18  & 4.66  & 0.00  & 1.38  & 40.00 & 21.00 & 18.40 & 7.31  & 56.68 & 28.80 & 4.12 \\
    Netherland (NL) & 2.26  & 45.20 & 18.80 & 1568.99 & 14.35 & 36.62 & 22.00 & 3.89  & 0.9   & 0.096 & 1.12  & 0.16  & 22    & 42.93 & 32.51 & 35.05 & 17.46 & 72.66 & 0.08  & 5.91  & 9.58  & 3.65  & 17.50 & 5.48  & 48.47 & 76.92 & 10.81 \\
    Austria (AT) & 1.90  & 39.70 & 14.90 & 1335.89 & 11.70 & 27.03 & 12.00 & 3.21  & 0.89  & 0.051 & 2.18  & 0.47  & 37    & 40.71 & 46.06 & 34.97 & 20.48 & 57.60 & 0.04  & 4.95  & 12.91 & 7.10  & 14.60 & 2.90  & 57.56 & 44.41 & 11.98 \\
    Poland (PL) & 0.63  & 43.50 & 3.70  & 276.71 & 5.04  & 1.92  & 11.00 & 1.64  & 0.54  & 0.029 & 0.47  & 1.24  & 12    & 13.27 & 11.39 & 8.34  & 3.50  & 3.68  & 0.02  & 0.58  & 5.25  & 5.90  & 10.00 & 5.54  & 49.44 & 39.59 & 6.45 \\
    Portugal (PT) & 1.90  & 35.00 & 9.60  & 873.39 & 9.02  & 21.23 & 25.00 & 1.99  & 0.66  & 0.069 & 0.6   & 0.64  & 23    & 42.08 & 37.81 & 25.59 & 7.75  & 6.65  & 0.01  & 0.70  & 7.21  & 4.47  & 10.90 & 3.74  & 36.70 & 44.40 & 6.27 \\
    Romania (RO) & 1.45  & 24.80 & 1.20  & 182.49 & 5.12  & 2.29  & 13.00 & 1.47  & 0.28  & 0.013 & 0.21  & 0.23  & 5     & 4.92  & 8.84  & 4.54  & 1.78  & 2.26  & 0.03  & 0.26  & 2.37  & 0.81  & 7.20  & 2.79  & 52.78 & 44.67 & 6.51 \\
    Slovenia (SI) & 3.55  & 43.00 & 11.60 & 1128.29 & 8.61  & 8.51  & 16.00 & 2.12  & 0.53  & 0.007 & 1.69  & 0.81  & 27    & 28.67 & 33.19 & 26.07 & 13.15 & 41.20 & 0.05  & 3.00  & 10.21 & 2.97  & 13.70 & 2.95  & 56.00 & 34.77 & 12.44 \\
    Slovakia (SK) & 2.25  & 33.40 & 2.90  & 407.83 & 5.50  & 9.13  & 9.00  & 1.41  & 0.85  & 0.008 & 0.33  & 0.58  & 20    & 16.72 & 22.44 & 13.94 & 8.41  & 9.96  & 0.04  & 0.45  & 4.30  & 1.06  & 10.00 & 7.37  & 66.48 & 34.80 & 19.12 \\
    Finland (FI) & 2.88  & 40.70 & 26.40 & 1576.00 & 10.82 & 19.88 & 26.00 & 5.98  & 0.95  & 0.107 & 1.94  & 0.32  & 34    & 44.10 & 37.26 & 38.32 & 16.77 & 63.05 & 0.05  & 8.29  & 11.44 & 4.50  & 15.70 & 2.84  & 44.64 & 62.40 & 9.27 \\
    Sweden (SE) & 2.91  & 47.30 & 29.60 & 1938.78 & 11.73 & 32.69 & 32.00 & 8.20  & 0.99  & 0.081 & 2.27  & 1.12  & 25    & 40.41 & 35.10 & 35.08 & 13.51 & 88.74 & 0.04  & 9.58  & 10.75 & 4.71  & 18.40 & 5.98  & 54.72 & 75.16 & 6.89 \\
    United Kingdom (UK) & 3.03  & 47.20 & 14.40 & 1151.26 & 14.49 & 42.95 & 10.00 & 3.34  & 0.56  & 0.103 & 1.12  & 0.67  & 28    & 32.58 & 45.45 & 19.03 & 24.70 & 43.19 & 0.03  & 3.25  & 7.32  & 3.03  & 18.40 & 6.94  & 54.75 & 82.90 & 20.81 \\
    \bottomrule
    \end{tabular}%

%% file: table_norm_performance_horizontal.tex
    \begin{tabular}{lccccccccccccccccccccccccccc}
    \toprule
              & \multicolumn{8}{c}{FC \;$\left(g_{1}\right)$} & \multicolumn{5}{c}{IN \;$\left(g_{2}\right)$} & \multicolumn{9}{c}{IA $\left(g_{(3)}\right)$} & \multicolumn{5}{c}{IMP $\left(g_{(4)}\right)$} \\
        \midrule
              & \multicolumn{3}{c}{HR \;$\left(g_{(1,1)}\right)$} & \multicolumn{3}{c}{ARS \;$\left(g_{(1,2)}\right)$} & \multicolumn{2}{c}{IFE \;$\left(g_{(1,3)}\right)$} & \multicolumn{3}{c}{FS \;$\left(g_{(2,1)}\right)$} & \multicolumn{2}{c}{FI \;$\left(g_{(2,2)}\right)$} & \multicolumn{3}{c}{FS \;$\left(g_{(3,1)}\right)$} & \multicolumn{3}{c}{LIN \;$\left(g_{(3,2)}\right)$} & \multicolumn{3}{c}{IAS \;$\left(g_{(3,3)}\right)$} & \multicolumn{2}{c}{EI \;$\left(g_{(4,1)}\right)$} & \multicolumn{3}{c}{FS \;$\left(g_{(4,2)}\right)$} \\
        \midrule
              & $g_{(1,1,1)}$   & $g_{(1,1,2)}$   & $g_{(1,1,3)}$   & $g_{(1,2,1)}$  & $g_{(1,2,2)}$  & $g_{(1,2,3)}$  & $g_{(1,3,1)}$   & $g_{(1,3,2)}$   & $g_{(2,1,1)}$   & $g_{(2,1,2)}$   & $g_{(2,2,1)}$   & $g_{(2,2,2)}$   & $g_{(2,2,3)}$   & $g_{(3,1,1)}$   & $g_{(3,1,2)}$   & $g_{(3,1,3)}$   & $g_{(3,2,1)}$ & $g_{(3,2,2)}$  & $g_{(3,2,3)}$ & $g_{(3,3,1)}$ & $g_{(3,3,2)}$  &$g_{(3,3,3)}$  & $g_{(4,1,1)}$  & $g_{(4,1,2)}$   & $g_{(4,2,1)}$   & $g_{(4,2,2)}$  & \multicolumn{1}{l}{$g_{(4,2,3)}$} \\
         \midrule     

	Belgium (BE) & 0.50  & 0.58  & 0.42  & 0.65  & 0.68  & 0.71  & 0.67  & 0.38  & 0.53  & 0.58  & 0.70  & 0.45  & 0.76  & 0.76  & 0.68  & 0.73  & 0.95  & 0.67  & 0.72  & 0.55  & 0.44  & 0.43  & 0.56  & 0.31  & 0.49  & 0.62  & 0.41 \\
    Bulgaria (BG) & 0.44  & 0.34  & 0.31  & 0.28  & 0.24  & 0.36  & 0.39  & 0.34  & 0.23  & 0.31  & 0.43  & 0.53  & 0.23  & 0.27  & 0.27  & 0.27  & 0.29  & 0.36  & 0.36  & 0.37  & 0.47  & 0.60  & 0.33  & 0.64  & 0.22  & 0.39  & 0.30 \\
    Czech Republic (CZ) & 0.48  & 0.34  & 0.46  & 0.43  & 0.39  & 0.45  & 0.39  & 0.46  & 0.67  & 0.30  & 0.52  & 0.61  & 0.51  & 0.51  & 0.41  & 0.54  & 0.47  & 0.41  & 0.46  & 0.40  & 0.40  & 0.43  & 0.44  & 0.54  & 0.72  & 0.40  & 0.66 \\
    Denmark (DK) & 0.79  & 0.60  & 0.87  & 0.90  & 0.72  & 0.61  & 0.84  & 1.00  & 0.85  & 0.52  & 0.73  & 0.34  & 0.63  & 0.57  & 0.61  & 0.55  & 0.55  & 1.00  & 0.40  & 0.73  & 0.51  & 0.64  & 0.59  & 0.48  & 0.48  & 0.67  & 0.38 \\
    Germany (DE) & 0.71  & 0.30  & 0.45  & 0.46  & 0.62  & 0.40  & 0.43  & 0.48  & 0.70  & 0.47  & 0.75  & 0.74  & 0.65  & 0.66  & 0.73  & 0.70  & 0.47  & 0.59  & 1.00  & 0.75  & 0.47  & 0.57  & 0.54  & 0.50  & 0.77  & 0.67  & 0.61 \\
    Estonia (EE) & 0.36  & 0.52  & 0.61  & 0.54  & 0.44  & 0.39  & 0.43  & 0.52  & 0.60  & 0.87  & 0.43  & 0.57  & 0.33  & 0.32  & 0.27  & 0.35  & 0.49  & 0.36  & 0.49  & 0.42  & 0.56  & 0.47  & 0.44  & 0.36  & 0.40  & 0.43  & 0.51 \\
    Ireland (IE) & 0.64  & 0.73  & 0.40  & 0.59  & 0.66  & 0.53  & 0.50  & 0.44  & 0.28  & 0.64  & 0.53  & 0.42  & 0.67  & 0.72  & 0.77  & 0.76  & 0.57  & 0.48  & 0.32  & 0.49  & 0.41  & 0.36  & 0.78  & 0.88  & 0.55  & 0.83  & 0.78 \\
    Grece (EL) & 0.37  & 0.51  & 0.35  & 0.40  & 0.49  & 0.50  & 0.22  & 0.35  & 0.49  & 0.25  & 0.33  & 0.54  & 0.37  & 0.56  & 0.61  & 0.60  & 0.59  & 0.40  & 0.52  & 0.36  & 0.39  & 0.37  & 0.41  & 0.50  & 0.10  & 0.42  & 0.59 \\
    Spain (ES) & 0.52  & 0.51  & 0.47  & 0.44  & 0.53  & 0.43  & 0.60  & 0.39  & 0.45  & 0.44  & 0.42  & 0.37  & 0.53  & 0.34  & 0.41  & 0.32  & 0.38  & 0.41  & 0.48  & 0.43  & 0.47  & 0.44  & 0.42  & 0.41  & 0.47  & 0.41  & 0.71 \\
    France (FR) & 0.48  & 0.57  & 0.67  & 0.44  & 0.61  & 0.69  & 0.37  & 0.66  & 0.57  & 0.63  & 0.62  & 0.43  & 0.47  & 0.58  & 0.63  & 0.60  & 0.55  & 0.52  & 0.49  & 0.60  & 0.41  & 0.44  & 0.51  & 0.47  & 0.64  & 0.61  & 0.67 \\
    Croatia (HR) & 0.46  & 0.35  & 0.33  & 0.36  & 0.26  & 0.35  & 0.31  & 0.34  & 0.35  & 0.49  & 0.37  & 0.72  & 0.51  & 0.43  & 0.48  & 0.43  & 0.38  & 0.38  & 0.48  & 0.37  & 0.39  & 0.35  & 0.39  & 0.35  & 0.33  & 0.21  & 0.31 \\
    Italy (IT) & 0.45  & 0.20  & 0.44  & 0.40  & 0.56  & 0.44  & 0.29  & 0.47  & 0.44  & 0.35  & 0.44  & 0.46  & 0.31  & 0.54  & 0.53  & 0.58  & 0.38  & 0.43  & 0.35  & 0.47  & 0.45  & 0.58  & 0.49  & 0.33  & 0.54  & 0.47  & 0.49 \\
    Cyprus (CY) & 0.25  & 0.82  & 0.41  & 0.57  & 0.56  & 0.42  & 0.24  & 0.42  & 0.26  & 0.57  & 0.27  & 0.31  & 0.51  & 0.54  & 0.49  & 0.58  & 0.51  & 0.39  & 0.28  & 0.38  & 0.98  & 0.45  & 0.61  & 0.16  & 0.41  & 0.62  & 0.29 \\
    Latvia (LV) & 0.32  & 0.53  & 0.42  & 0.30  & 0.24  & 0.40  & 0.65  & 0.53  & 0.38  & 0.70  & 0.29  & 0.46  & 0.31  & 0.24  & 0.32  & 0.26  & 0.28  & 0.36  & 0.59  & 0.35  & 0.43  & 0.39  & 0.36  & 0.52  & 0.28  & 0.49  & 0.32 \\
    Lithuania (LT) & 0.37  & 0.80  & 0.39  & 0.34  & 0.23  & 0.35  & 0.63  & 0.43  & 0.58  & 0.62  & 0.32  & 1.00  & 0.27  & 0.55  & 0.39  & 0.58  & 0.60  & 0.36  & 0.83  & 0.38  & 0.42  & 0.38  & 0.29  & 0.45  & 0.28  & 0.22  & 0.44 \\
    Luxembourg (LU) & 0.34  & 0.73  & 0.63  & 0.75  & 0.63  & 1.00  & 0.63  & 0.65  & 0.50  & 0.46  & 0.42  & 0.28  & 0.65  & 0.60  & 0.80  & 0.61  & 0.45  & 0.40  & 0.33  & 0.45  & 0.94  & 0.82  & 0.92  & 0.47  & 0.54  & 0.81  & 0.37 \\
    Hungary (HU) & 0.33  & 0.29  & 0.40  & 0.36  & 0.35  & 0.38  & 0.43  & 0.41  & 0.30  & 0.50  & 0.51  & 0.53  & 0.39  & 0.29  & 0.27  & 0.28  & 0.37  & 0.48  & 0.45  & 0.41  & 0.38  & 0.36  & 0.41  & 0.77  & 0.80  & 0.44  & 0.58 \\
    Malta (MT) & 0.24  & 0.37  & 0.43  & 0.39  & 0.52  & 0.43  & 0.43  & 0.50  & 0.32  & 0.24  & 0.35  & 0.37  & 0.53  & 0.45  & 0.48  & 0.48  & 0.32  & 0.38  & 0.29  & 0.42  & 0.96  & 1.00  & 0.71  & 0.75  & 0.61  & 0.29  & 0.28 \\
    Netherland (NL) & 0.59  & 0.60  & 0.67  & 0.70  & 0.77  & 0.66  & 0.65  & 0.56  & 0.68  & 0.69  & 0.54  & 0.29  & 0.51  & 0.68  & 0.51  & 0.65  & 0.66  & 0.73  & 0.79  & 0.72  & 0.47  & 0.47  & 0.67  & 0.58  & 0.48  & 0.69  & 0.52 \\
    Austria (AT) & 0.52  & 0.49  & 0.59  & 0.63  & 0.63  & 0.57  & 0.43  & 0.51  & 0.67  & 0.48  & 0.81  & 0.42  & 0.82  & 0.65  & 0.69  & 0.65  & 0.74  & 0.66  & 0.53  & 0.65  & 0.53  & 0.61  & 0.53  & 0.35  & 0.62  & 0.42  & 0.56 \\
    Poland (PL) & 0.27  & 0.56  & 0.34  & 0.31  & 0.29  & 0.33  & 0.41  & 0.39  & 0.43  & 0.38  & 0.37  & 0.73  & 0.31  & 0.26  & 0.22  & 0.23  & 0.30  & 0.37  & 0.39  & 0.36  & 0.40  & 0.56  & 0.31  & 0.59  & 0.50  & 0.38  & 0.36 \\
    Portugal (PT) & 0.52  & 0.39  & 0.47  & 0.49  & 0.49  & 0.51  & 0.71  & 0.41  & 0.51  & 0.56  & 0.41  & 0.49  & 0.53  & 0.67  & 0.58  & 0.50  & 0.41  & 0.39  & 0.34  & 0.37  & 0.44  & 0.50  & 0.35  & 0.43  & 0.31  & 0.42  & 0.36 \\
    Romania (RO) & 0.43  & 0.18  & 0.29  & 0.28  & 0.29  & 0.34  & 0.46  & 0.38  & 0.25  & 0.30  & 0.31  & 0.32  & 0.17  & 0.14  & 0.19  & 0.17  & 0.25  & 0.37  & 0.47  & 0.34  & 0.36  & 0.35  & 0.17  & 0.34  & 0.55  & 0.42  & 0.37 \\
    Slovenia (SI) & 0.85  & 0.55  & 0.52  & 0.57  & 0.47  & 0.39  & 0.52  & 0.42  & 0.42  & 0.28  & 0.68  & 0.56  & 0.61  & 0.48  & 0.51  & 0.51  & 0.55  & 0.57  & 0.58  & 0.52  & 0.48  & 0.44  & 0.48  & 0.36  & 0.60  & 0.34  & 0.58 \\
    Slovakia (SK) & 0.59  & 0.36  & 0.33  & 0.35  & 0.31  & 0.40  & 0.37  & 0.37  & 0.65  & 0.28  & 0.34  & 0.46  & 0.47  & 0.31  & 0.37  & 0.32  & 0.43  & 0.41  & 0.50  & 0.36  & 0.39  & 0.36  & 0.31  & 0.75  & 0.75  & 0.34  & 0.82 \\
    Finland (FI) & 0.72  & 0.51  & 0.84  & 0.70  & 0.59  & 0.50  & 0.73  & 0.71  & 0.71  & 0.74  & 0.75  & 0.36  & 0.76  & 0.70  & 0.57  & 0.71  & 0.64  & 0.68  & 0.56  & 0.87  & 0.50  & 0.50  & 0.58  & 0.35  & 0.43  & 0.57  & 0.47 \\
    Sweden (SE) & 0.72  & 0.64  & 0.91  & 0.81  & 0.64  & 0.62  & 0.86  & 0.88  & 0.74  & 0.62  & 0.83  & 0.68  & 0.57  & 0.65  & 0.54  & 0.66  & 0.56  & 0.82  & 0.51  & 0.96  & 0.49  & 0.51  & 0.71  & 0.63  & 0.58  & 0.68  & 0.38 \\
    United Kingdom (UK) & 0.75  & 0.64  & 0.58  & 0.57  & 0.78  & 0.71  & 0.39  & 0.52  & 0.44  & 0.72  & 0.54  & 0.50  & 0.63  & 0.54  & 0.68  & 0.40  & 0.85  & 0.58  & 0.47  & 0.54  & 0.44  & 0.44  & 0.71  & 0.72  & 0.58  & 0.74  & 0.88 \\
\bottomrule   
\end{tabular}%